%% file: Convex.tex
\documentclass[]{article}  
\usepackage{url}
\usepackage{graphicx}
\usepackage{amsfonts}
\usepackage{amssymb}
\usepackage{latexsym}

\input{mac}

\def\g{{\gamma}}
\def\t{{\theta}}
\def\d{{\delta}}
\def\e{{\varepsilon}}

\newcommand{\notyet}[1]{}

\usepackage{xcolor}

\begin{document}

\title{Partitioning Regular Polygons into Circular Pieces I:\\
Convex Partitions}

\author{%
Mirela Damian%
   \thanks{Department of Computer Science, Villanova University, Villanova,
PA 19085, USA.
   \protect\url{mirela.damian@villanova.edu}.}
\and
  Joseph O'Rourke%
    \thanks{Department of Computer Science, Smith College, Northampton, MA
      01063, USA.
      \protect\url{orourke@cs.smith.edu}.
       Supported by NSF Distinguished Teaching Scholars award
       DUE-0123154.}
}

\maketitle

\begin{abstract}
We explore an instance of the question of partitioning a polygon
into pieces, each of which is as ``circular'' as possible, in the sense of
having an aspect ratio close to $1$.
The \emph{aspect ratio} of a polygon is the ratio of
the diameters of the smallest circumscribing circle to the largest
inscribed disk.
The problem is rich even for partitioning regular polygons into convex pieces,
the focus of this paper.
We show that the optimal (most circular) partition for an equilateral
triangle has an infinite number of pieces, with the lower bound
approachable to any accuracy desired by a particular finite partition.
For pentagons and all regular $k$-gons, $k > 5$, the unpartitioned
polygon is already optimal.  The square presents an interesting
intermediate case.  Here the one-piece partition is not optimal,
but nor is the trivial lower bound approachable.  We narrow the
optimal ratio to an aspect-ratio gap of $0.01082$ with several somewhat intricate partitions.
\end{abstract}

\section{Introduction}
At the open-problem session of the 14th Canadian Conference on Computational
Geometry,\footnote{
   Lethbridge, Alberta, Canada, August 2002.} 
the first author posed the question of finding 
a polynomial-time algorithm for partitioning a polygon into
pieces, each with an aspect ratio no more than a given $\g > 1$~\cite{do-op02-03}.
The \emph{aspect ratio} of a polygon $P$ is the ratio of
the diameters of the smallest circumscribing circle to the largest
inscribed \emph{indisk}.
(We will use ``circumcircle'' and ``indisk'' to emphasize that
the former may overlap but the latter cannot.)
If the pieces of the partition must have their vertices chosen among
$P$'s vertices, i.e, if  ``Steiner points'' are disallowed,
then a polynomial-time algorithm
is known~\cite{d-eaaco-02}.
Here we explore the question without this restriction, but
with two other restrictions:  the pieces are all convex,
and the polygon $P$ is a regular $k$-gon.
Although the latter may seem highly specialized, in fact many of
the issues for partitioning an arbitrary polygon arise already with
regular polygons.  The specialization to convex pieces is
both natural, and most in concert with the applications mentioned below.
Partitions employing nonconvex pieces will be explored in~\cite{do-prpcp2-03}.
Our emphasis in this paper is not on algorithms, but on the partitions
themselves.

\subsection{Notation}
A \emph{partition} of a polygon $P$ is a collection of polygonal \emph{pieces}
$P_1, P_2, \ldots$ such that
$P = \cup_i P_i$ and no pair of pieces share an interior point.
To agree with the packing and covering literature, we will use $\g$ for
the aspect ratio, modified by subscripts and superscripts as appropriate.
$\g_1(P)$ is the one-piece $\g$: the ratio of the radius of the smallest
circumcircle of $P$, to the radius of the largest disk inscribed in $P$.
$\g(P)$ is the maximum of all the $\g_1(P_i)$ for all pieces $P_i$ in
a partition of $P$; so this is dependent upon the particular
partition under discussion.  
Both the partition and the argument ``$(P)$'' will often be dropped
when clear from the context.
$\g^*(P)$ is the minimum $\g(P)$ over all convex partitions of $P$.
Our goal is to find $\g^*(P)$ for the regular $k$-gons.

To present our results, we introduce one more bit of notation,
forward-referencing Section~\secref{One-angle.Lower.Bound}:
$\g_\t$ is the ``one-angle lower bound,'' a lower bound
derived from one angle of the polygon, ignoring all else.  This
presents a trivial lower bound on any partition's aspect ratio.

\subsection{Table of Results}
Our results are summarized in Table~\tabref{Results}.
It makes sense that regular polygons for large $k$ are as circular
as one could get.  We prove that this holds true for all
$k \ge 5$.
For 
an equilateral
triangle, the optimal partition can approach but never achieve
the lower bound $\g_{60^\circ}=3/2$, for finite partitions.
The square is 
an interesting
intermediate case.  Here the one-piece partition is not optimal,
but nor is the trivial lower bound approachable.
So $\g^* \in ( \g_{90^\circ}, \g_1 ) $.
We narrow the
optimal ratio to a small gap by raising the lower bound and lowering
the upper bound.

\begin{table}[htbp]
\begin{center}
\begin{tabular}{| l || c | c || c | c |}
	\hline

\emph{Regular Polygon}
	& $\g_1$
	& $\g_\t$
	& $\g^*$
        & $k^*$

       	\\ \hline \hline
Triangle
	& $2.00000$
	& $1.50000$
	& $\g_\t$
        & $\infty$

       	\\ \hline
Square
	& $1.41421$
	& $1.20711$
	& $\in [1.28868, 1.29950]$ 
        & $\infty$?

       	\\ \hline
Pentagon
	& $1.23607$
	& $1.11803$
	& $\g_1$
	& $1$

       	\\ \hline
Hexagon
	& $1.15470$
	& $1.07735$
	& $\g_1$
	& $1$

       	\\ \hline
Heptagon
	& $1.10992$
	& $1.05496$
	& $\g_1$
	& $1$

       	\\ \hline
Octagon
	& $1.08239$
	& $1.04120$
	& $\g_1$
	& $1$

       	\\ \hline \hline
$k$-gon
	& $1/\cos ( \pi/k )$
	& $[1 + \csc( \t/2 )]/2$
	& \multicolumn{2}{c |}{\mbox{}}

        \\ \hline
\end{tabular}
\caption{Table of Results on Regular Polygons.
$\g_1$: one-piece partition;
$\g_\t$: single-angle lower bound; $\t$: angle at corner;
$\g^*$: optimal partition; $k^*$: number of pieces in optimal partition.
}
\end{center}
\tablab{Results}
\end{table}

\subsection{Motivation}
\seclab{Motivation}
Partitions of polygons into components that satisfy various shape
criteria have been the focus of considerable research. 
Algorithms have been developed~\cite{k-85} 
that produce
partitions of simple polygons into the fewest convex polygons,
spiral polygons, star-shaped polygons, or monotone polygons,
when all the vertices of the partition pieces are selected 
from among the polygon vertices, i.e., Steiner points are not employed.
Permitting Steiner points
makes optimal partitions much more difficult to find.
A notable success here is the polynomial-time
algorithm of Chazelle and Dobkin~\cite{cd-ocd-85},
which partitions a simple polygon into the fewest number
of convex pieces.
See Keil~\cite{k-00} and Bern~\cite{b-97} for surveys
of polygon partitioning.

Our motivation for investigating circular partitions is 
for their advantages in several application areas,
for circular polygons have a number of desirable properties. 
Because they can be tightly circumscribed by a circle,
they support quick collision detection tests, important in
motion planning, dynamics simulations, and virtual reality.
In graphics scenes containing a number of nonintersecting circular 
polygons, only a few might intersect a region small relative
to the size of the polygons.
This property has been used~\cite{of-94} to reduce the 
time complexity of a particular motion planning algorithm.
Examples of other applications for which
circular objects confer an advantage are range searching~\cite{of-96},
simulation of physically-based motion~\cite{khmsz-96}, and
ray tracing~\cite{fd-96}.
Partitions into circular pieces are relevant to all
these applications. 

Our focus on convex pieces is motivated by computational 
geometry applications that have simple and efficient and solutions if the 
input consists of convex pieces only.
For instance, algorithms for collision detection, radiosity calculations, 
shading, and clipping run an order of magnitude faster
on convex polygons than similar algorithms designed to handle
arbitrary polygons.
Convex polygons are also preferable in computer graphics because they are
easy to represent, manipulate, and render.

There is an interesting connection between the problem studied in
this paper and packings and coverings.  For any partition of $P$, the
collection of indisks for each piece of the partition forms a packing
of $P$ by disks, and the collection of circumcircles enclosing each
piece form a covering of $P$.  We have found this connection to packing
and covering more relevant when the pieces are not restricted
to be convex~\cite{do-prpcp2-03}, but even for convex pieces it is
at least suggestive.  
For example, 
the notorious difficulty of packing equal disks in a square
may relate to the apparent difficulty of our problem 
for the square, which leads to
a packing of the square with unequal disks.
In any case, despite the connections, we have not found in the literature
any work that directly addresses our particular problem.

\subsection{Outline of Paper}
Rather than follow the ordering in the table, we proceed in
order of increasing difficulty: starting with
the easiest result (pentagons and $k > 5$, Section~\secref{Pentagon}),
moving next to the equilateral triangle (Section~\secref{EquiTri}),
and finally to the square (Section~\secref{Square}).
A preliminary section (Section~\secref{One.One}) establishes some simple lemmas used throughout,
and we look to the natural next steps for this work in a
final Discussion, Section~\secref{Discussion}.

\section{One-piece Partition and One-angle Lower Bound}
First we establish two simple lemmas used throughout the rest of the paper.
\seclab{One.One}
\subsection{One-piece Partition}
\seclab{One-piece.Partition}
\begin{figure}[htbp]
\centering
\includegraphics[width=0.3\linewidth]{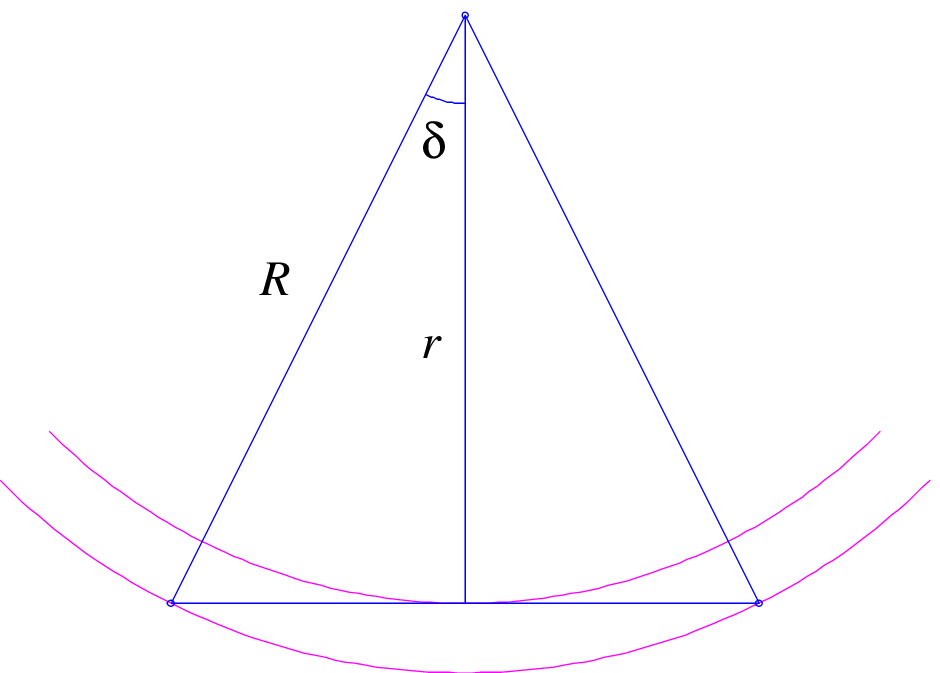}
\caption{Aspect Ratio of Regular Polygons.}
\figlab{one.piece}
\end{figure}
\begin{lemma} {\em \bf (Regular Polygon).}
The aspect ratio $\g_1$ of a regular $k$-gon is 
 \[ \g_{1} =  \frac{1}{\cos(\pi/k)} \]
\lemlab{Reg.Poly}
\end{lemma}
\begin{pf}
Both the indisk and the circumcircle are
centered on the polygon's centroid.
Referring to Figure~\figref{one.piece}, the angle $\delta$
formed by circumradius $R$ and inradius $r$ is $\pi/k$. It follows immediately
that $R/r = 1/\cos(\pi/k)$. 
\end{pf}

\subsection{One-angle Lower Bound}
\seclab{One-angle.Lower.Bound}
\begin{figure}[htbp]
\centering
\includegraphics[width=0.2\linewidth]{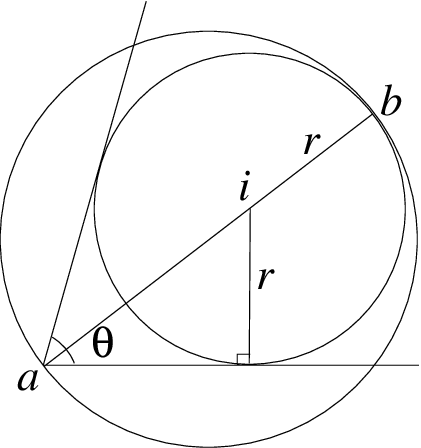}
\caption{One-angle lower bound.}
\figlab{lower.bound}
\end{figure}

\begin{lemma} {\em \bf (One-Angle Lower Bound).}
If a polygon $P$ contains a convex vertex of internal angle $\t$, then the 
aspect ratio of a convex partition of $P$ is no smaller than
$\g_{\t}$, with
 \[ \g_{\t} =  \frac{1 + \csc(\t/2)}{2} \]
\lemlab{one.angle.lower.bound}
\end{lemma}
\begin{pf}
Either the whole angle $\t$
is included in one piece of the partition, or it is split and shared
among several pieces. The latter gives an even lower bound.  So consider
the former.

Then the indisk must fit in the angle and the circumcircle 
must include the 
corner and surround the indisk; see Figure~\figref{lower.bound}.
We get a lower bound by maximizing the 
inradius and minimizing the circumradius.  This happens when the indisk 
and the circumcircle are tangent where they intersect the bisector of 
the angle $\t$. 
Referring to Figure~\figref{lower.bound}, the inradius is $|ib| = r$, 
the circumcircle's diameter is $|ab| = 2R$ and 
\[ \sin(\t/2) = \frac{r}{2R - r}\]
From this we get $\g_{\t} = R / r = (1+\csc(\t/2)/2$, which concludes
the lemma.  
\end{pf}

\noindent
Note that  $\g_{\t}$ cannot be achieved by a polygonal
partition, a point to which we return in Lemma~\lemref{not.lower.bound} below.

\section{Pentagon}
\seclab{Pentagon}
It is clear that for large enough $k$, $\g^* = \g_1$ for a $k$-gon.
The only question is for which $k$ does this effect take over.
The answer is $k=5$:

\begin{theorem}
For a regular pentagon ($k=5$), the optimal convex partition is
just the pentagon itself, i.e., $\g^* = \g_1$ and $k^* = 1$.
\theolab{pentagon}
\end{theorem}
\begin{pf}
The regular pentagon $P$ has $\g_1(P)= \sqrt{5}-1 \approx 1.23607$ 
(by the Regular
Polygon Lemma~\lemref{Reg.Poly}).  
This is the $\g$-value for competitive partitions to
beat.
\begin{enumerate}
\item 
Suppose some piece has an edge incident to a vertex of the
pentagon, and so splits the $108^\circ$ angle there. 
Then some piece has an
angle $\le 54^\circ$.
Lemma~\lemref{one.angle.lower.bound} then yields 
$\g_{54^\circ} \approx 1.60134$.
Because this is larger than $1.23607$, no such partition
could improve over $\g_1$.

\item 
So assume there is a partition, none of whose edges are incident
to a vertex of $P$.  We consider three cases for corner pieces:
two acute, one acute, or both obtuse angles adjacent to $v$.

\begin{enumerate}
\item 
Suppose some piece of the partition that includes a vertex $v$ of $P$
contains two acute or right ($\le 90^\circ$) angles adjacent to $v$.  
Then this piece already exceeds
$\g_1(P)$, as is evident from 
Figure~\figref{Pentagon.proof}(a).
\begin{figure}[htbp]
\centering
\begin{tabular}{c@{\hspace{0.2in}}c@{\hspace{0.2in}}c}
\includegraphics[width=0.28\linewidth]{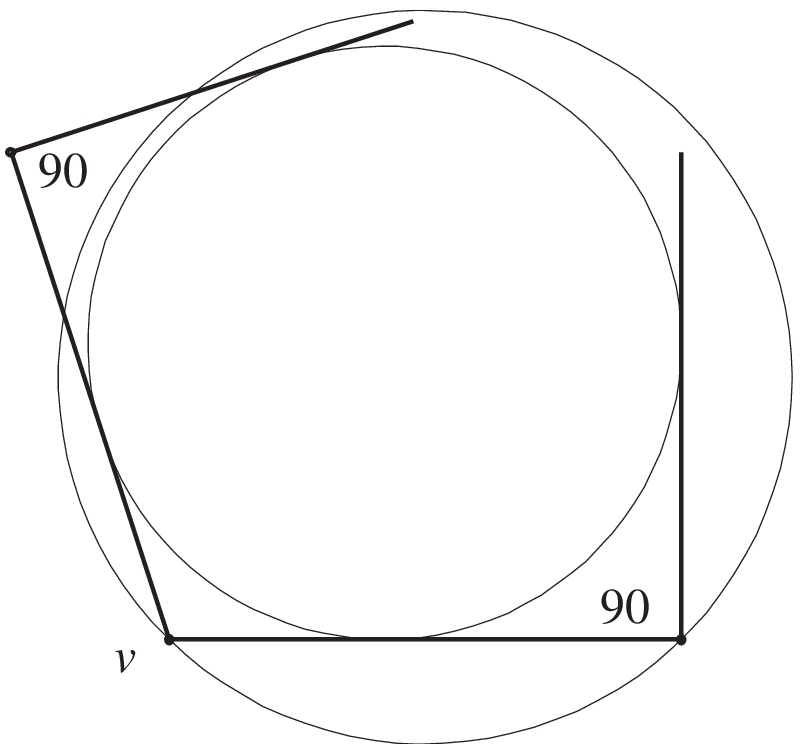} &
\includegraphics[width=0.28\linewidth]{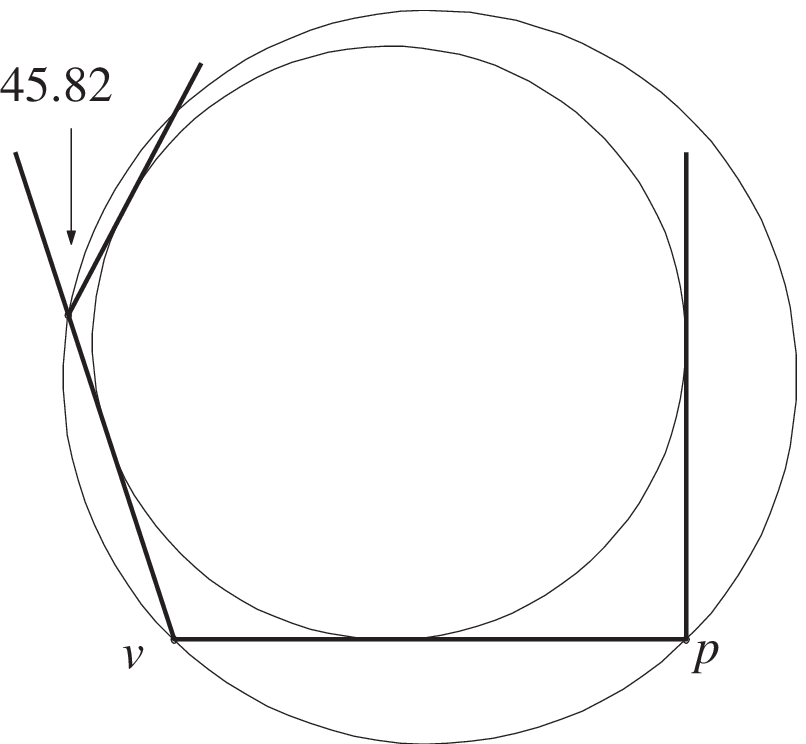} &
\includegraphics[width=0.26\linewidth]{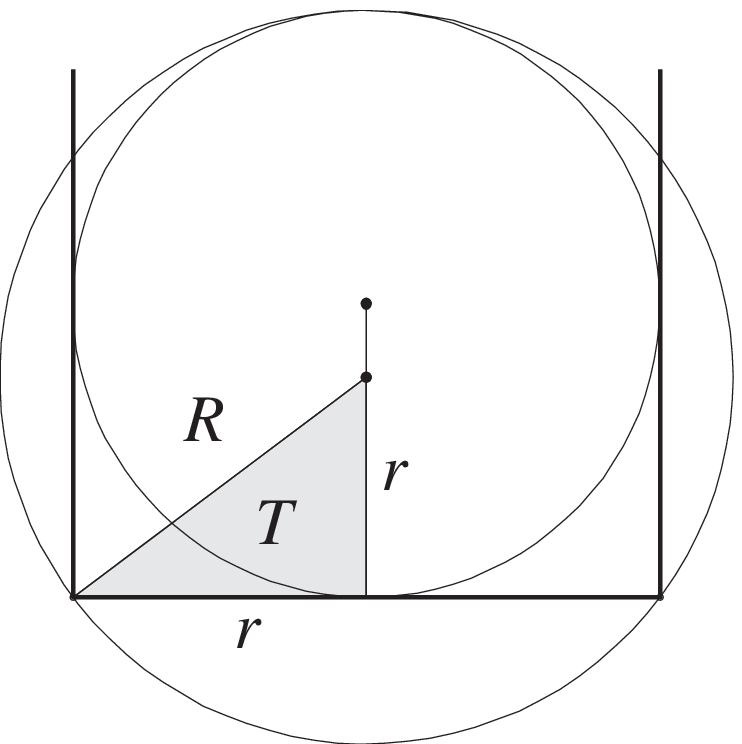} \\
(a) & (b) & (c)
\end{tabular}
\caption{Optimal pentagon partition (a) $\g_1(P) \approx 1.23607$ is 
insufficient to cover a corner piece with two right or acute angles 
adjacent to $v$.
(b) $\g_1(P) \approx 1.23607$ creates an angle $\le 45.8^\circ$ 
in a piece adjacent to a corner piece with one acute and one obtuse angle
(c) $\g_1 > 1.25$ for a piece with two adjacent acute angles.}
\figlab{Pentagon.proof}
\end{figure}
Therefore every piece that encompasses a vertex $v$ of $P$ has at most one
angle $\le 90^\circ$ adjacent to $v$.

\item 
Suppose some piece of the partition that includes a vertex $v$ of
$P$ contains one acute or right angle adjacent to $v$.  Note that
here the one-angle lower bound Lemma~\lemref{one.angle.lower.bound} 
does not suffice to settle this case, as that
yields $\g_{90^{\circ}} = (1+\sqrt{2})/2 \approx 1.20711 < \g_1(P)$. 
Suppose, to be generous, that this one
nonobtuse angle is as large as possible, $90^\circ$. Then if the exterior
circle is to include the foot $p$ of that perpendicular and $v$, then the
adjacent piece must have an acute angle $\le 45.8^\circ$ to the other side of
$v$, as simple geometric computations show; see
Figure~\figref{Pentagon.proof}b. 
Applying Lemma~\lemref{one.angle.lower.bound} 
yields $\g_{45.8^{\circ}} = 1.78$, which well exceeds $\g_1(P)$.  Thus that adjacent
piece is inferior.

\item 
Suppose every piece including a vertex $v$ of $P$ has only obtuse
angles adjacent to $v$.  
Then no two
pieces containing adjacent vertices of $P$ can themselves be adjacent,
for otherwise at their join along the side $s$ of $P$, an angle would
be obtuse on both sides.  So
there must be one or more intervening pieces. 
It is easy to see that, for each side $s$ of $P$, one
of these intervening pieces must have acute angles at both its corners on
$s$. 
For if the angle pairs alternate acute/obtuse, ..., acute/obtuse, one of the corner
pieces will have an acute angle on $s$. 
Such a piece must have $\g$ greater than the aspect ratio of a piece
with right angles at both its corners on $s$.
From the right triangle $T$ in Figure~\figref{Pentagon.proof}c we 
get $(2r-R)^2 + r^2 = R^2$, which yields 
$R/r = 1.25 > \g_1(P)$.  
\end{enumerate}

Having exhausted all cases, we are left with the conclusion that no
partition can improve upon $\g_1(P)$.  Therefore $k^* = 1$.

\end{enumerate}
\end{pf}


\section{Equilateral Triangle}
\seclab{EquiTri}
Recall from Table~\tabref{Results}
that the one-angle lower bound for the equilateral triangle
is $\g=3/2$.
In this section we show that this lower bound can be approached,
but never achieved, by a finite convex partition.
That it cannot be achieved is a general result:
\begin{lemma}
For any polygon $P$, the one-angle lower bound $\g_{\t}(P)$ can
never be achieved by a finite convex partition.
\lemlab{not.lower.bound}
\end{lemma}
\begin{pf}
Recall from Figure~\figref{lower.bound}
that $\g_{\t}$ is determined by a circumcircle $C_1$
tangent to an indisk $C_0$ nestled in the minimum angle corner
of $P$.
Let $x$ be the point of tangency between these circles.
As is clear from Figure~\figref{tangent.smooth},
any partition piece $Q$ covering $x$ and the corner must either lie
exterior to $C_1$ or interior to $C_0$ in a neighborhood of $x$.
Either possibility forces $\g(Q) > \g_{\t}(P)$.
\end{pf}
\begin{figure}[htbp]
\centering
\includegraphics[width=0.6\linewidth]{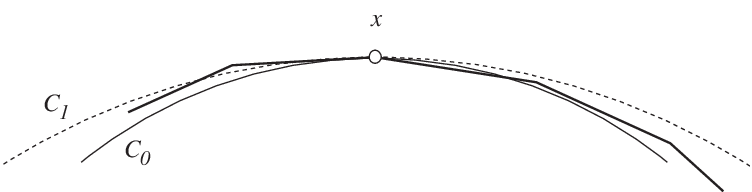}
\caption{A polygonal approximation in the vicinity
of the point of tangency $x$ between the indisk $C_0$ and the 
circumcircle $C_1$ must either go exterior to $C_1$
(left) or interior to $C_0$ (right).}
\figlab{tangent.smooth}
\end{figure}

Our proof that the lower bound of $\g=3/2$ can be approached
for an equilateral triangle follows five steps:
\begin{enumerate}
\item  $\g=3/2$ suffices to partition any rectangle 
(Lemma~\lemref{rect.gamma}).
\item  $\g=3/2$ suffices to partition any ``$80^\circ$-quadrilateral''
(Lemma~\lemref{80deg.quad}).
\item  $\g=3/2$ suffices to approximately partition the region between
any pair of  ``$80^\circ$-curves''
(Lemma~\lemref{80deg.curve}).
\item The corners of an equilateral triangle can be covered by
a polygon with $\g_1$ as near to $3/2$ as desired
(Figure~\figref{inter.80}a).
\item The remaining ``interstice'' can be partitioned
into regions bound by $80^\circ$-curves
(Figure~\figref{inter.80}b).
\end{enumerate}

\subsection{Rectangles}
We first establish that $\g=3/2$ permits partitioning
of any rectangle.  In fact, we prove that $\g > \sqrt{2}$ suffices.

Define the \emph{aspect ratio} $a(R)$ of a rectangle $R$ to be
ratio of the length of its longer side to that of its shorter side.
We start by noting that $\g=3/2$ allows one-piece partitions
for some aspect ratios strictly greater than $1$.
\begin{lemma}
Any rectangle $R$ with $a(R) < \frac{3}{2 \sqrt{2}} \approx 1.06066$
has $\g_1 < 3/2$.
\lemlab{1.5}
\end{lemma}

\noindent
See Figure~\figref{circ.1.5} where $r = 1$, $R =
\frac{\sqrt{17}}{2\sqrt{2}}$ and therefore $R/r < 3/2$. 
\begin{figure}[htbp]
\centering
\includegraphics[width=0.3\linewidth]{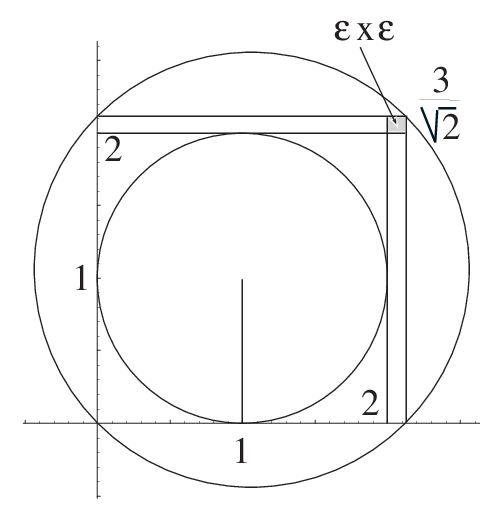}
\caption{The range of rectangles fitting between the
in-circle and out-circle have $\g_1 < 3/2$.}
\figlab{circ.1.5}
\end{figure}

Now we generalize to arbitrary aspect ratios, and
return to $\g=3/2$ below.

\begin{lemma}
Any rectangle $R$ may be partitioned into a finite number of
rectangles, each with an aspect ratio of no more than $1+\e$, for
any $\e > 0$.
\lemlab{rect.epsilon}
\end{lemma}
\begin{pf}
Let $R$ have dimensions $a \times b$, with $a \le b$.
Define 
\begin{eqnarray}
k_b &=& \lceil 1/\e \rceil \\
k_a &=& \lceil a / (b/k_b) \rceil
\end{eqnarray}
We want to cut up the long dimension into $k_b$ squares.
Because $k_b > 1/\e$, the portion left over 
can be absorbed by expanding each square by $\e$ in that direction,
for we know that $\e k_b > 1$.
In other words, the left-over
portion needs less than one more square to cover it, and
the $\e$'s suffice to give us this one more.
We then arrange to slice the $a$ dimension into equal-sized strips
whose width $\d=a/k_a$ assures the desired $\g^* = 1+\e$. See
Figure~\figref{rect.epsilon}.

\end{pf}

\noindent
\begin{figure}[htbp]
\centering
\includegraphics[width=0.4\linewidth]{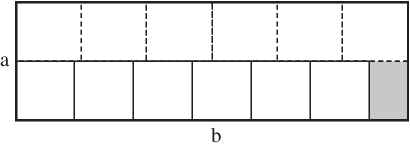}
\caption{$k_b=6$ and $k_a=2$; $\e=0.15$. 
The bottom row shows  $\d \times \d$ squares,
leaving a gap less than $\d$, and the top row show $\d(1+\e_1) \times \d$ squares,
with $\e_1 < \e$,
that fully fill the strip.}
\figlab{rect.epsilon}
\end{figure}

\begin{lemma}
Any rectangle $R$ may be partitioned into a finite number pieces with 
$\g(R) = \g$, for any $\g > \sqrt{2}$.
\lemlab{rect.gamma}
\end{lemma}
\begin{pf}
Recall that $\g_1$ for a square is $\sqrt{2}$.
So any $\g > \sqrt{2}$ leaves a bit of slack, permitting rectangles
of aspect ratio $1+\e$, for some $\e > 0$,  to have that $\g$. 
For example, Lemma~\lemref{1.5} shows that for $\g=3/2$,
$\e = \frac{3}{2 \sqrt{2}} - 1 \approx 0.06066$.
Now apply Lemma~\lemref{rect.epsilon}.
\end{pf}

\noindent
The number of pieces in the implied partition depends on $\e$, which depends
on $\g$.  For $\g=3/2$, each strip will have $k_b = \lceil 1/0.06066 \rceil = 17$
rectangles.  Of course it is likely that much more efficient partitions
are possible.

\subsection{$80^\circ$-Quadrilaterals}
We next extend the rectangle lemma to slightly skewed quadrilaterals.

\begin{lemma}
There are quadrilaterals with one corner a right angle, and its
opposite corner any angle within $90^\circ \pm 11^\circ$,
which have $\g_1 < 3/2$.
\lemlab{80deg.quad}
\end{lemma}
\begin{pf}
Any quadrilateral fitting between the indisk and the circumcircle of 
Figure~\figref{circ.1.5} satisfies $\g_1 < 3/2$. 
See Figure~\figref{circ.80} for an example.
\begin{figure}[htbp]
\centering
\includegraphics[width=0.35\linewidth]{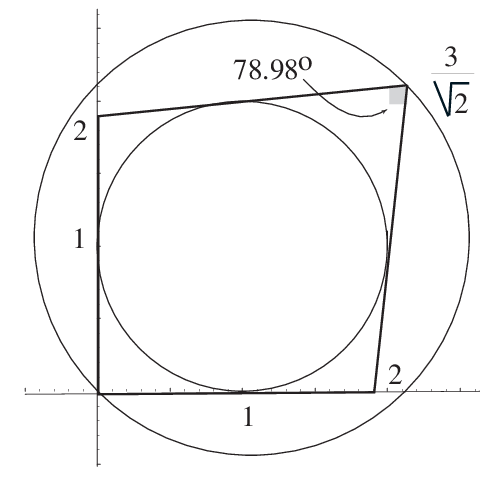}
\caption{The angle indicated is
$2 \cos^{-1} \frac{9}{2 \sqrt{34}} \approx 90^\circ - 11.02^\circ$.
The shaded square is the $\e \times \e$ square in Figure~\protect\figref{circ.1.5}.
}
\figlab{circ.80}
\end{figure}
\end{pf}

\noindent
Define an \emph{$80^\circ$-quadrilateral} as one 
whose corner angles fall within
$90^\circ \pm 11^\circ$.

\begin{lemma}
Any $80^\circ$-quadrilateral $Q$ 
may be partitioned with $\g(Q)=3/2$.
\lemlab{quad}
\end{lemma}
\begin{pf}
Partition $Q$ into four ``quarters'' with orthogonal lines $L_1$ and $L_2$
through the centroid.
We partition each quarter $Q_i$ separately.
Start at the exterior corner of $Q_i$, and create the largest
quadrilateral satisfying Lemma~\lemref{80deg.quad}, with the quadrilateral's right
angle flush against a quartering line, say $L_1$.
See piece $A$ in Figure~\figref{quad.80}.
Continue in this manner, with each piece flush against $L_1$,
until the next piece would overlap into the
next quarter ($B$ and $C$ in the figure).
Switch strategies to take a smaller piece flush with $L_2$ and the boundary
of $Q_i$ ($D$ in the figure).
Finally, we are left with a rectangle; cover according to Lemma~\lemref{rect.epsilon}.
\end{pf}
\begin{figure}[htbp]
\centering
\includegraphics[width=0.6\linewidth]{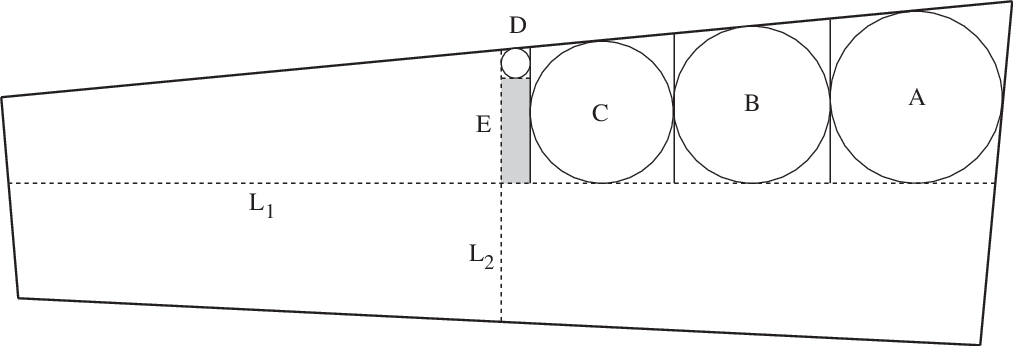}
\caption{Partitioning one quarter of a quadrilateral.}
\figlab{quad.80}
\end{figure}

\subsection{$80^\circ$-Curves}
Define a \emph{pair of $80^\circ$-curves} 
$A(t)$ and $B(t)$ as two curves parametrized by
$t \in [0,1]$,
that satisfy the property that each segment $C(t)$
whose endpoints are $A(t)$ and $B(t)$, 
meets the curves at angles in the range $90^\circ \pm 11^\circ$.

\begin{lemma}
Let $A(t)$ and $B(t)$ be a pair of
$80^\circ$-curves.
Then the region bounded by $A$, $B$, $C(0)$, and $C(1)$ may
be approximated to any given degree of accuracy by a region
$R$ that can be finitely partitioned with $\g(R) \le 3/2$.
\lemlab{80deg.curve}
\end{lemma}
\begin{pf}
Choose an $n > 1$ and partition the range of $t$ into $n$ equal
intervals.  The quadrilaterals 
$$(\; A(i/n), B(i/n), B((i+1)/n), A((i+1)/n) \;)$$
for $i=0,\ldots,n-1$,
satisfy Lemma~\lemref{80deg.quad} by hypothesis, and so each
can individually be partitioned with $\g \le 3/2$.
Clearly as $n \rightarrow \infty$, the approximation of
$R$ to the region bounded by the curves improves in accuracy.
\end{pf}

\subsection{Partition of Interstice}
The overall plan of the partition of an equilateral triangle
is to first cover each $60^\circ$-corner with a large convex polygon
that approximates the lower bound.
Figure~\figref{inter.80}a shows an example that is only $0.00120$
above $\g=3/2$.

Three such pieces leave an ``interstice'' $I$ whose exact shape depends
on how large the corner pieces are. 
In the center of the interstice (centered on the
triangle centroid) we place a regular hexagon.
Then the goal is to cover the remainder of the interstice,
between the corner pieces and the hexagon, with $\g=3/2$.
We will employ $80^\circ$-curves near the hexagon; closer
to the triangle side curves will not be necessary.
The most difficult part of the construction is joining the curves 
to the hexagon.
As 
Figure~\figref{inter.80}b illustrates,
we partition the remaining
$240^\circ$ exterior to each hexagon vertex into three $80^\circ$ angles.
These provide termination points for the $80^\circ$-curves.
\begin{figure}[htbp]
\centering
\begin{tabular}{c@{\hspace{1in}}c}
\includegraphics[width=0.35\linewidth]{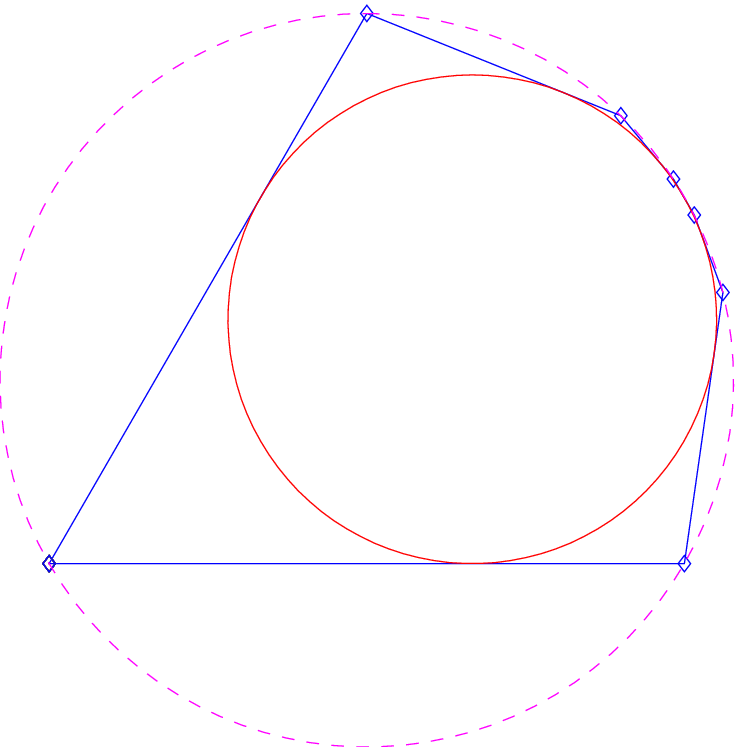} &
\includegraphics[width=0.3\linewidth]{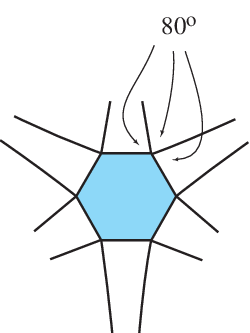} \\
(a) & (b)
\end{tabular}
\caption{(a) A heptagon covering a $60^\circ$-corner, with $\g=1.5012$.
(b) The $120^\circ$ hexagon angle meets three $80^\circ$ angles.}
\figlab{inter.80}
\end{figure}

The next task is to obtain an explicit form of the most critical
pair of $80^\circ$-curves, between
the corner indisk $C_0$ and one of the curves incident
to a hexagon corner.
We choose to make this second curve an arc of a circle, although
there are many choices.
The construction starts with a circle $C_a$ and two radii
$\pm 10^\circ$ from the horizontal.  These radii bound a circular
arc $a$ of angle $20^\circ$, which meets the base of the hexagon
in an $80^\circ$ angle, and extends down to line parallel to
the base at another $80^\circ$ angle.
See Figure~\figref{arcs.circs}a.
The arc $a$ meets one of our criteria:  it is incident to the hexagon
at an $80^\circ$ angle.  Now we consider $a$ paired with an arc
of $C_0$.  Note that any segment $s$ incident to $a$ from the left, which
starts within the shaded region $R$, forms an angle $\ge 80^\circ$
with $a$, because $R$ is delimited by rays $\pm 10^\circ$ from 
the endpoints of $a$.

\begin{figure}[htbp]
\centering
\includegraphics[width=0.55\linewidth]{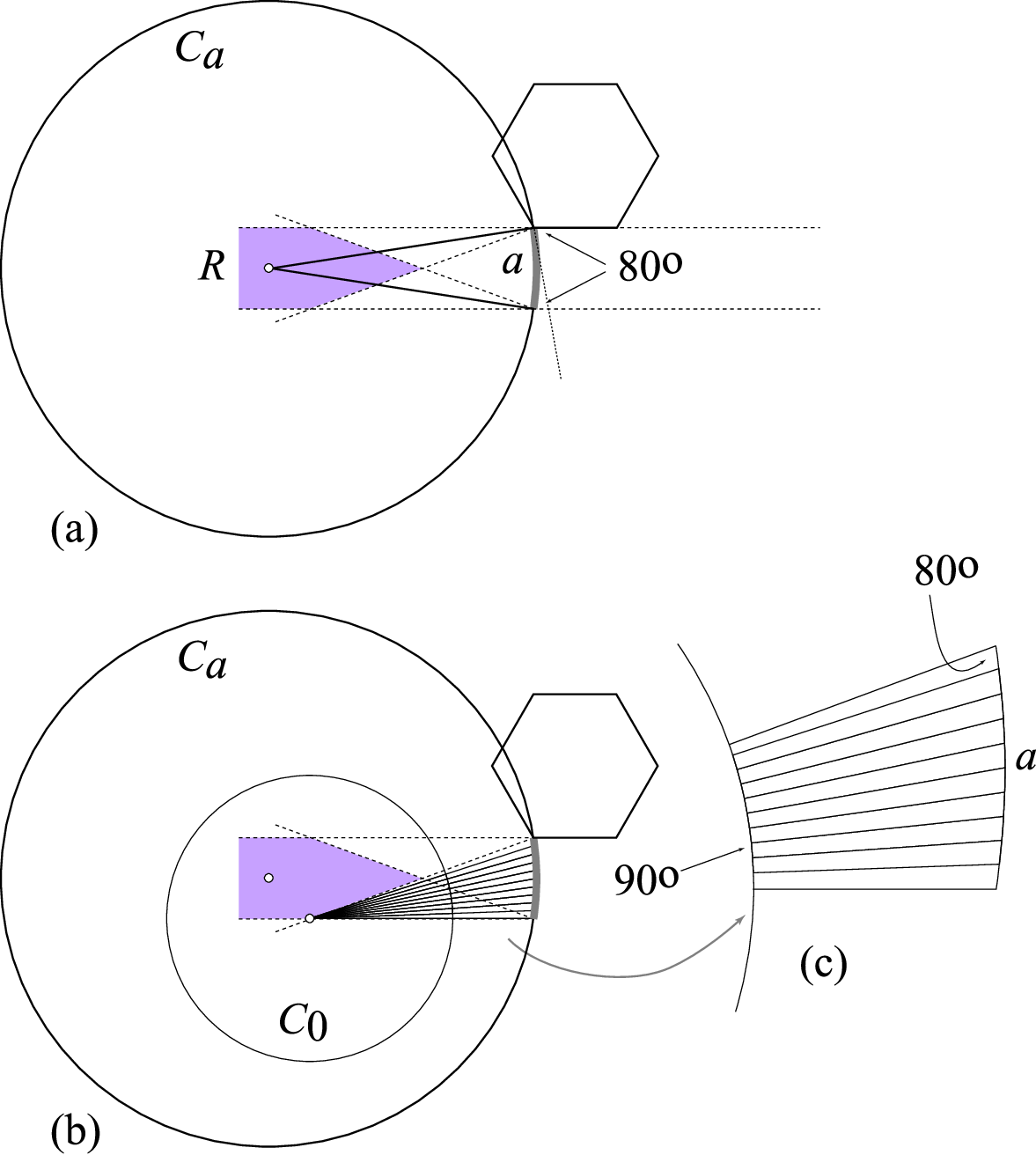}
\caption{(a) $C_a$ determines arc $a$; $R$ is delimited by rays 
 $\pm 10^\circ$ from the endpoints of $a$. 
(b) $C_0$ is centered on a vertex of $R$.
(c) A magnified view of the quadrilaterals determined by the pair of
$80^\circ$ circle arcs.}
\figlab{arcs.circs}
\end{figure}

Now consider placing the center $C_0$ on the vertex
of $R$ indicated in Figure~\figref{arcs.circs}b.
We are going to partition the region between $C_0$ and $a$ with radial
rays from the center of $C_0$.  Because the center of $C_0$ is in $R$,
all these rays satisfy the $\ge 80^\circ$ limit where they hit $a$;
and of course they are orthogonal to $C_0$.  So the quadrilaterals
determined by these rays are $80^\circ$-quadrilaterals.
Finally, as shown in Figure~\figref{arcs.circs}c,
the angle of the uppermost quadrilateral incident to the hexagon
corner is exactly  $80^\circ$, because $C_0$ was centered on
a ray that achieves this.
Therefore, we have in fact achieved
the goal set out in Figure~\figref{inter.80}b: the exterior angle
around the hexagon vertex is partitioned into three  $80^\circ$ angles.

The overall design of the full partition is illustrated in
Figure~\figref{eqtri.arcs}.
\begin{figure}[htbp]
\centering
\includegraphics[width=0.55\linewidth]{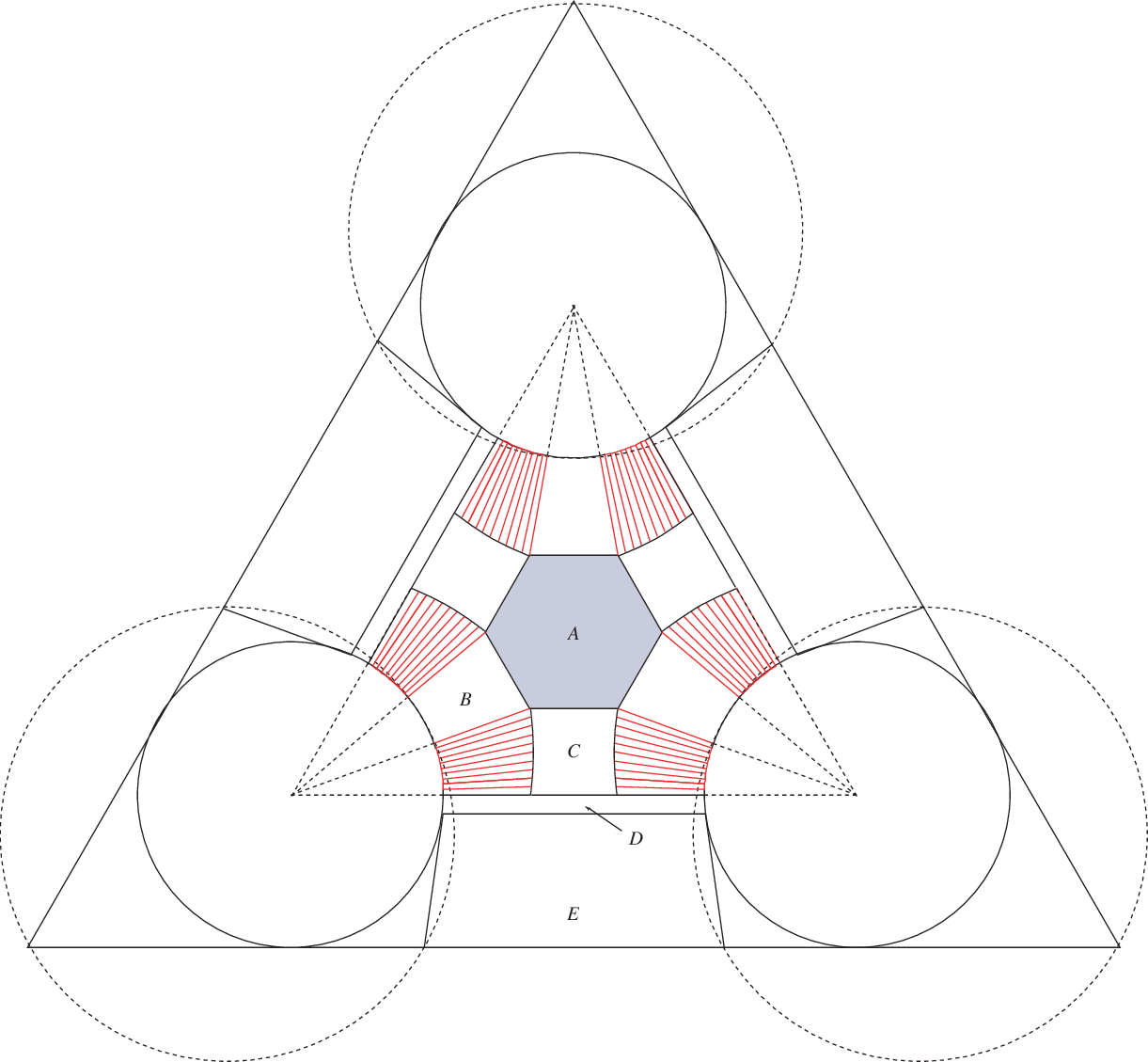} 
\caption{Overall design of partition of equilateral triangle}
\figlab{eqtri.arcs}
\end{figure}
This figure shows several parts of the construction, but is not
a full partition.  The region $B$ is easily partitioned into
$80^\circ$-quadrilaterals by more radial rays from the center
of $C_0$.  How many rays depends on the polygonal approximation to
the corner piece; see again Figure~\figref{inter.80}a.
The region $C$ can be partitioned by horizontal lines;
again all the quadrilaterals are $80^\circ$-quadrilaterals
(see the lower $80^\circ$ angle in Figure~\figref{arcs.circs}a).
The $80^\circ$-curves terminate on the line 
through the centers of the two bottom incirles.
Piece $D$ of the partition would again be partitioned 
by horizontal segments, as many as are needed to accomodate
the polygonal approximation to $C_0$ there.
Finally, the large remaining quadrilateral $E$ has base
angles $> 82^\circ$, and so is already an $80^\circ$-quadrilateral.
A possible partition into $80^\circ$-quadrilaterals is shown in 
Figure~\figref{eqtri.all}.
\begin{figure}[htbp]
\centering
\includegraphics[width=0.6\linewidth]{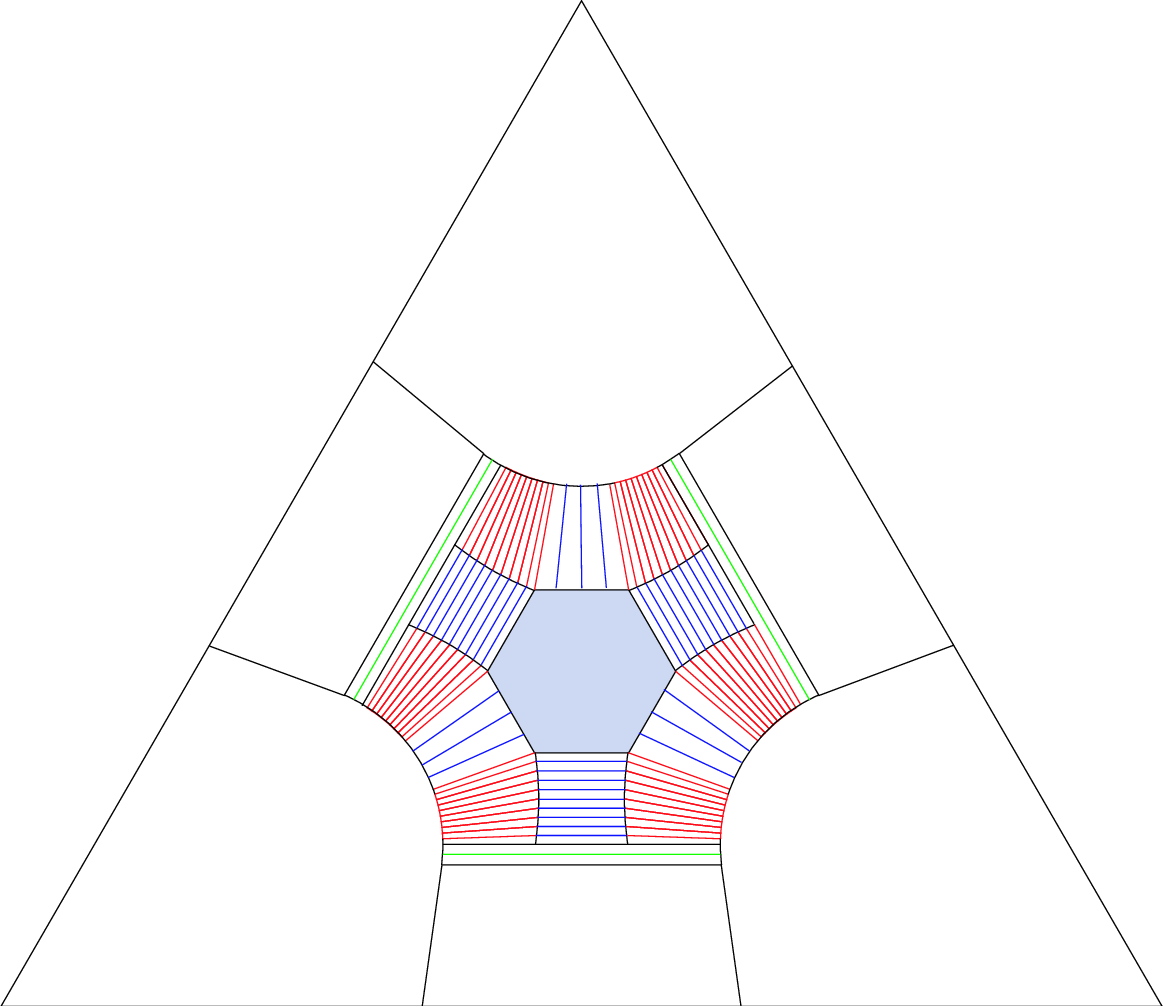}
\caption{An equilateral triangle partitioned into corner
pieces, a hexagon, and $80^\circ$-quadrilaterals .}
\figlab{eqtri.all}
\end{figure}

We have established our theorem:

\begin{theorem}
An equilateral triangle may be partitioned into a finite number of
pieces with ratio $\g$, for any $\g > 3/2$.
As $\g$ approaches $3/2$, the number of pieces goes to infinity.
\theolab{equi.tri}
\end{theorem}
\begin{pf}
The hexagon has $\g_1 = 1.15470 < 3/2$.
Partition each quadrilateral according to Lemma~\lemref{80deg.quad},
using $\g \le 3/2$.
The result is a partition whose $\g$ is determined by the corner pieces,
the only ones with $\g > 3/2$, but
which, as Figure~\figref{inter.80}a illustrates, can approach
the lower bound of $3/2$ as closely as desired.
\end{pf}

\section{Square}
\seclab{Square}
The square is intermediate between the equilateral triangle
and the regular pentagon in several senses:
(a) Unlike the pentagon, the one-piece partition $\g_1 =1.41421$
is not optimal;
(b) Unlike the equilateral triangle, the one-angle lower
bound of $\g_{90^\circ}=1.20711$ (Table~\tabref{Results})
cannot be approached.
Although we believe a result similar to 
Theorem~\theoref{equi.tri} holds---there is a lower bound
that can be approached but not reached for a finite partition---we
have only confined the optimal ratio
$\g^*$ to the range $[1.28868, 1.29950]$,
leaving a gap of $0.01082$.
We first show in Section~\secref{Pentagons.on.Side}
that the one-piece partition is not optimal
through a series of increasingly complex partitions.
Then we describe in Section~\secref{Hexagons.on.Side}
the example that establishes our upper
bound $\g^* \le 1.29550$.
Section~\secref{Square.Lower.Bound} presents the proof of the
lower bound, $\g^* \ge 1.28868$.
Finally, we close with a conjecture in Section~\secref{Square.Open}
and supporting evidence.

\subsection{Pentagons on Side}
\seclab{Pentagons.on.Side}
To improve upon the one-piece partition, pieces that have five
or more sides must be employed, for the square itself is the
most circular quadrilateral.
One quickly discovers that covering the side of the square is
challenging and crucial to the structure of the overall partition.
We first explore partitions that use pentagons around the square 
boundary.
Figure~\figref{Pentagonal} shows a $12$-piece partition, with a central octagon, that
already improves upon $\g_1 =1.41421$, achieving
$\g=1.33964$; but it is easily seen
to be suboptimal.
\begin{figure}[htbp]
\centering
\begin{tabular}{c@{\hspace{1.1in}}c}
\includegraphics[width=0.4\linewidth]{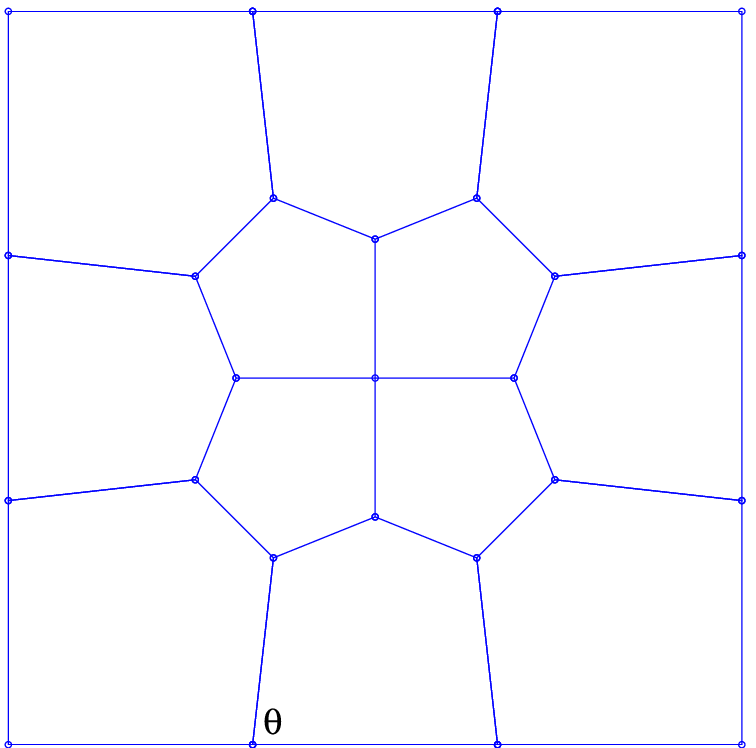} &
\includegraphics[width=0.4\linewidth]{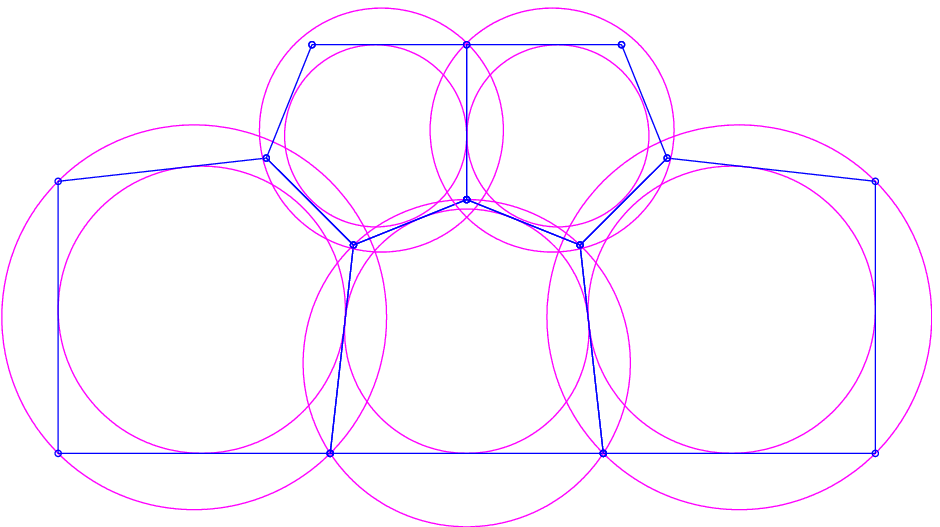} \\
(a) & (b)
\end{tabular}
\caption{$12$-piece partition achieving $\g=1.33964$ 
(a) Overall design. Here $\t = 89.62^\circ$. 
(b) Details of critical indisks and circumcircles.}
\figlab{Pentagonal}
\end{figure}
Our next attempt,
Figure~\figref{sq.4.convex},
is a $21$-piece partition improving
to $\g=1.32348$.  It's center is a near-regular $16$-gon.
\begin{figure}[htbp]
\centering
\begin{tabular}{cc}
\includegraphics[width=0.5\linewidth]{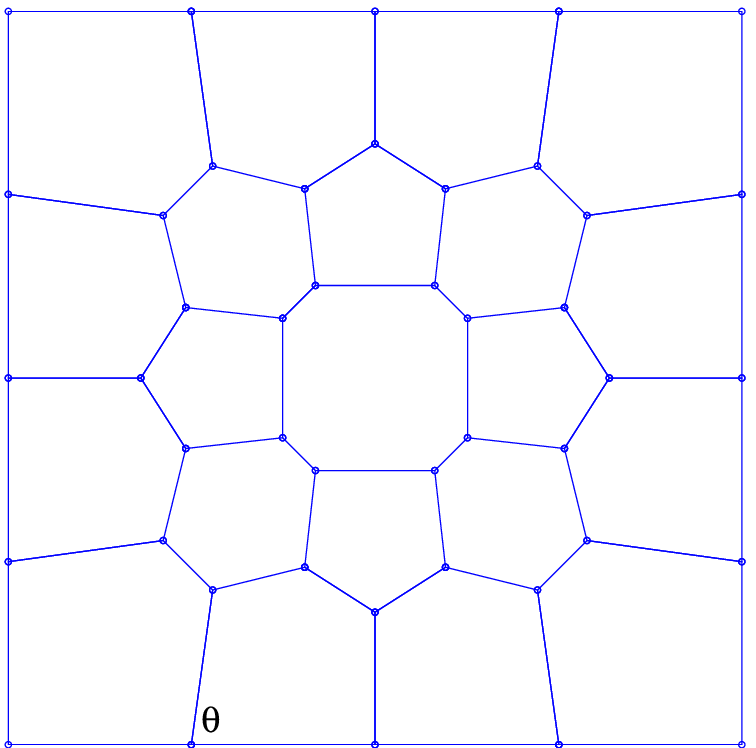} &
\includegraphics[width=0.5\linewidth]{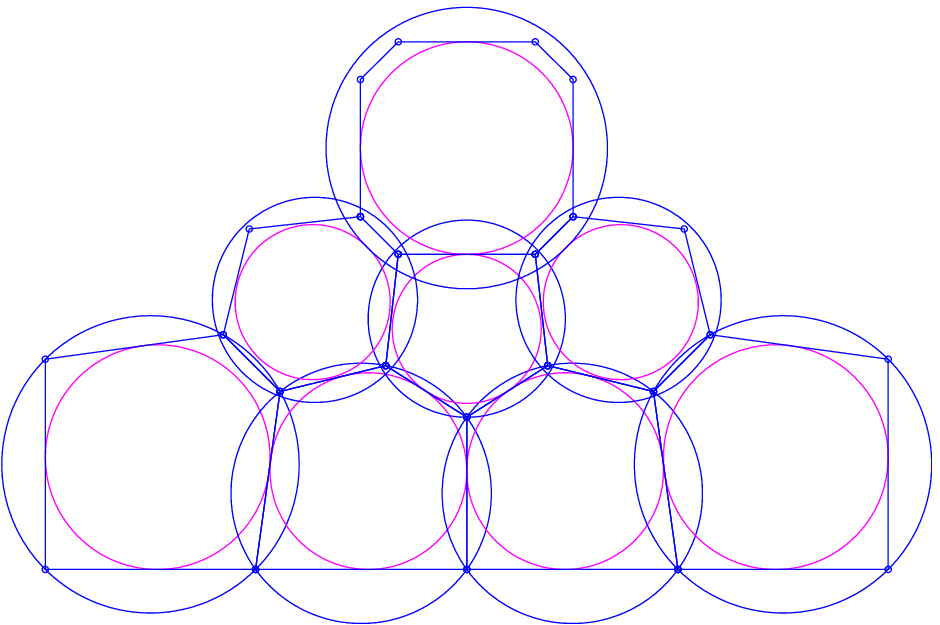} \\
(a) & (b)
\end{tabular}
\caption{$21$-piece partition achieving $\g=1.32348$. (a) Overall design.
Here $\t=82.16^\circ$. (b) Details of critical indisks and circumcircles.}
\figlab{sq.4.convex}
\end{figure}
Our most elaborate pentagon example is the $92$-piece partition
shown in Figure~\figref{sq.16-gon}, which achieves $\g=1.31539$.
It includes four heptagons adjacent to the corner pieces, and
four dodecagons at the center, with all remaining pieces hexagons or pentagons.
Note that throughout this series, the angle $\t$ on the square
side between the corner piece and its immediate neighbor 
is a critical angle.  One wants this small so that the corner
piece can be circular; but too small and the adjacent piece cannot
be circular.  It is exactly this tension that we will
exploit in Section~\secref{Square.Lower.Bound} to establish
a lower bound.
\begin{figure}[bhtp]
\centering
\begin{tabular}{cc}
\includegraphics[width=0.5\linewidth]{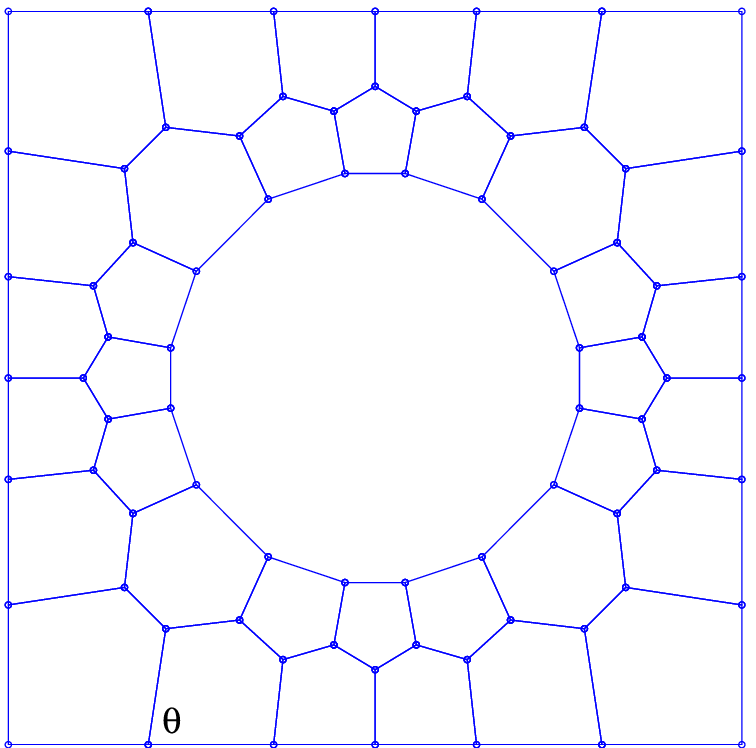} &
\includegraphics[width=0.5\linewidth]{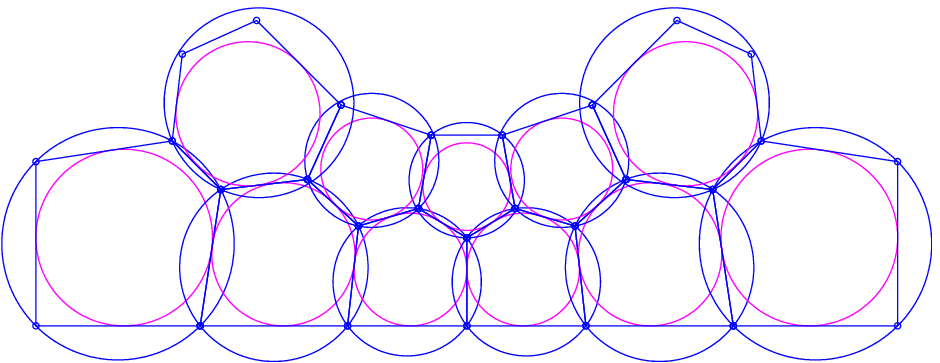}
\end{tabular} 
\caption{$37$-piece partition achieving $\g=1.31539$.
(a) Overall design. Here $\t = 81.41^\circ$. 
The central polygon is a near-regular $16$-gon.
(b) Details of critical indisks and circumcircles.}
\figlab{sq.16-gon}
\end{figure}

We will leave it as a claim without proof that any square partition
that covers the entire boundary of the square with pentagons
must have $\g \ge 1.31408$.
So Figure~\figref{sq.16-gon} cannot be much improved.

\subsection{Hexagons on Side}
\seclab{Hexagons.on.Side}
To make further advances, it is necessary to move beyond pentagons.
Figure~\figref{stainedglass} shows our best partition, which employs
four corner pentagons and four hexagons along each side of the square.
As is apparent from the figure, it is no longer straightforward
to fill the interior after covering the boundary.
\begin{figure}[htbp]
\centering
\includegraphics[width=0.61\linewidth]{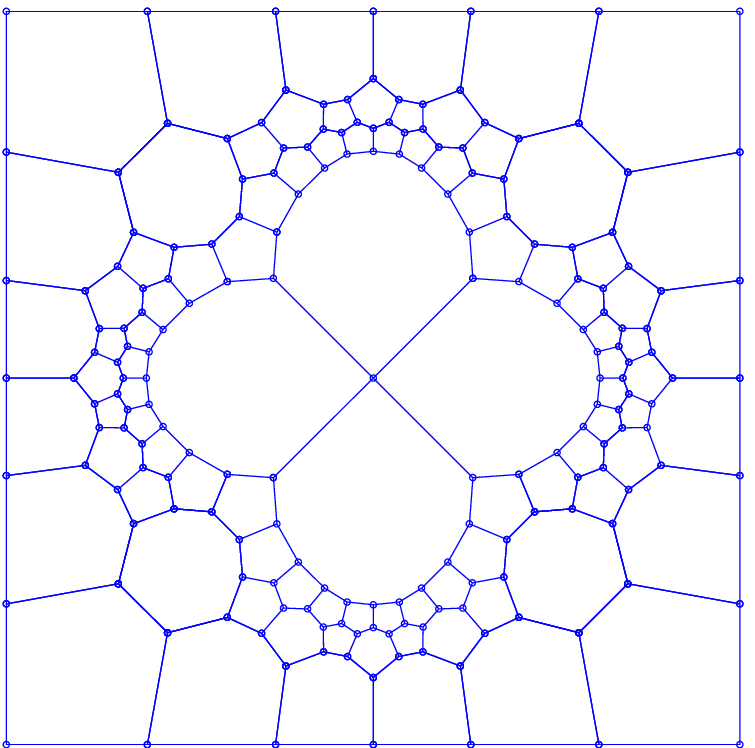} 
\caption{$92$-piece partition achieving $\g=1.29950$}
\figlab{stainedglass}
\end{figure}

Figure~\figref{stainedglass2} shows the indisks and circumcircles
for the pieces of the partition.  Only the central tear-drop shape
and the corner pieces on the ``second level'' have significant slack 
with respect to the achieved $\g=1.29950$.

\begin{figure}[htbp]
\centering
\includegraphics[width=0.65\linewidth]{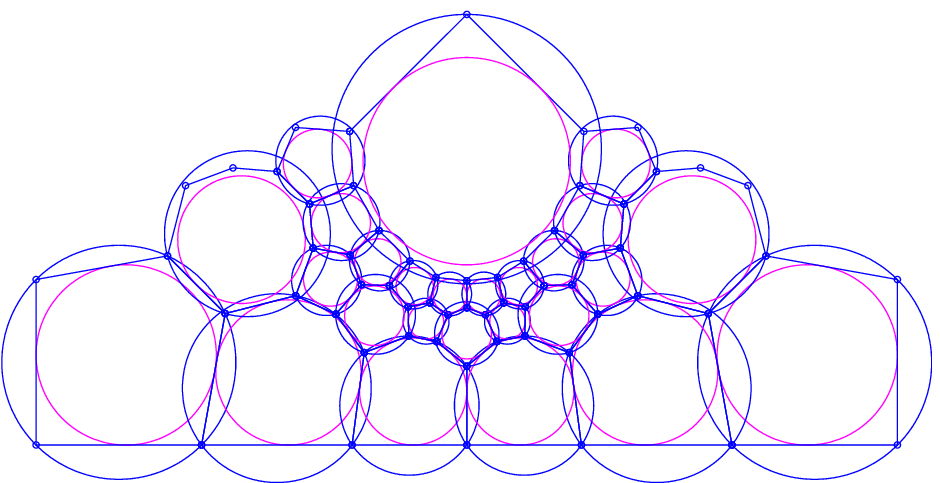} 
\caption{Details of critical indisks and circumcircles from Figure~\figref{stainedglass}}
\figlab{stainedglass2}
\end{figure}

We again leave it as claim without proof that further improvements
here will not be large: any square partition
that covers the entire boundary of the square with hexagons
must have $\g \ge 1.29625$.

\subsection{Lower Bound on $\g^*$}
\seclab{Square.Lower.Bound}
One quickly develops a sense that
the lower bound provided by Lemma~\lemref{one.angle.lower.bound},
$\g_{90^\circ} = (1 + \sqrt{2})/2 \approx 1.20711$,
is not approachable.
We derive here a larger lower bound, $\g^{*} \ge 1.28868$,
by focussing on the piece adjacent to a corner piece, which
we call a ``bottleneck piece.''

\begin{theorem}
The optimal aspect ratio for a convex partition of a square is at least
$$\g^* \ge  1.28868$$
\theolab{square.lower.bound}
\end{theorem}

\noindent
The proof is in three steps.
Let $a$, $b$, and $c$ be the first three piece vertices along the bottom square side,
with $a$ the square corner, as in 
Figure~\figref{sqlb0}.
\begin{figure}[htbp]
\centering
\includegraphics[width=0.5\linewidth]{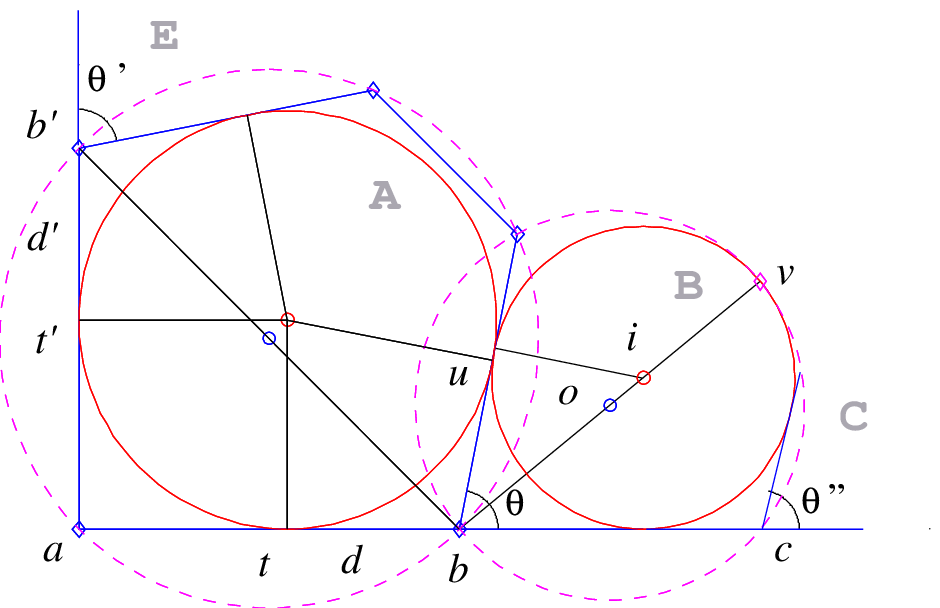}
\caption{Balancing aspect ratios for $A$ and $B$.}
\figlab{sqlb0}
\end{figure}
We first derive a lower bound by concentrating on $b$ only (Lemma~\lemref{sq.A}),
and then improve that by focussing on $c$ as well  (Lemma~\lemref{sq.B}).
Finally, we show that this bound applies to any partition  (Lemma~\lemref{sq.C}).

\begin{lemma}
Pieces $A$ and $B$ of Figure~\figref{sqlb0} (the corner piece and
the bottleneck piece) must satisfy
$$\min \{ \g_1(A), \g_1(B) \}  \ge 1.28782$$
\lemlab{sq.A}
\end{lemma}
\begin{pf}
Let the inscribed circle for the partition piece $A$ that covers the corner
have radius $r_A = 1$. We know that $\g_1(A) \ge \g_{90^\circ} \approx 1.20711$. 
We will show that in fact $\g_1(A) > 1.28868$. 

Let $B$ be the piece adjacent to $A$ along the bottom side of the square. The
angles of the two pieces at their shared square side point $b$ are
supplementary: the internal angle at $b$ in $A$ is $\pi - \t$. 
The lower bound $\g_{90^\circ}$ derives from the 
$90^\circ$ angle $a$ of $A$, but if $\t < 90^\circ$, then the lower bound for 
the piece $B$ is larger. 
So we compute two bounds on $\g$ as functions of $\t$: one
for piece $A$ and one for piece $B$. The overall optimum is achieved when these
two bounds are in balance. So we set them equal and solve for $\t$. We
then use $\t$ to compute $\g^{*}$.

Let $C$ and $E$ be the partition pieces adjacent to $B$ and $A$ along the bottom and
left side of the square respectively (see again Figure~\figref{sqlb0}).
Without loss of generality, we make two assumptions: 
\begin{itemize}
\item[(a)] the internal angle $\t$ of $B$ at point $b$ is no greater than 
the internal angle $\t'$ of $E$ at point $b'$; 
otherwise, we can balance the aspect ratios of
$A$ and $E$ to get a larger $\g^{*}$.
\item[(b)] the internal angle $\t$ of $B$ at point $b$ is no less that 
the internal angle $\t''$ of $C$ at point $c$; 
otherwise, we can balance the aspect 
ratios of $A$ and $C$ to get a larger $\g^{*}$.
\end{itemize}

First we derive $\g_1(A)$ as a function of $\t$. 
We must have $|tb| = |bu|$ and therefore the distance 
$d$ between points $t$ and $b$ in Figure~\figref{sqlb0} is: 
\begin{equation} 
d = 1 / \tan((\pi-\t)/2) = \tan(\t/2)
\label{eq:lb1}
\end{equation}
Similarly, the distance between $t'$ and $b'$ is $d' = \tan(\t'/2)$, or
\begin{equation}
d' \ge \tan(\t/2)
\label{eq:lb2}
\end{equation}
A's circumradius is $R_A = |bb'|/2 = \frac{1}{2}\sqrt{(1+d)^2 + (1+d')^2}$,
which together with (\ref{eq:lb1}) and (\ref{eq:lb2}) yields 
\begin{equation}
\g_1(A) = R_A \ge \frac{1+\tan(\t/2)}{\sqrt{2}}
\label{eq:ga}
\end{equation}

\noindent
For the piece $B$, the lower bound Lemma~\lemref{one.angle.lower.bound} 
yields:
\begin{equation}
\g_1(B) \ge \frac{1+\csc(\t/2)}{2}
\label{eq:gb}
\end{equation}
To minimize $\g$, we balance the bounds $\g_1(A)$ and 
$\g_1(B)$ and solve for theta. From this we get  
$\t \approx 78.79^{\circ}$ and $\g_1(A) = \g_1(B) \approx 1.28782$.
\end{pf}

\begin{figure}[htbp]
\centerline{
\begin{tabular}{cc}
\includegraphics[width=3in]{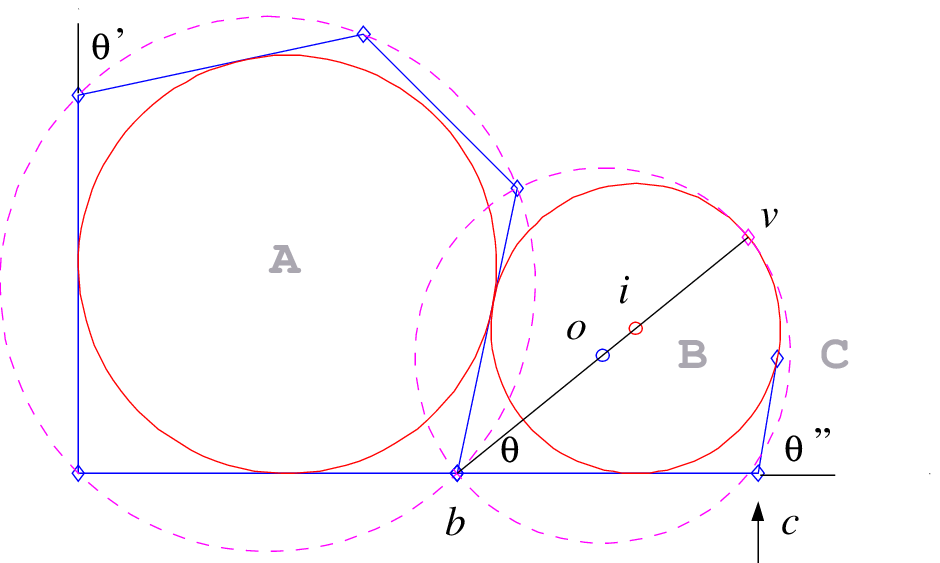} &
\includegraphics[width=3in]{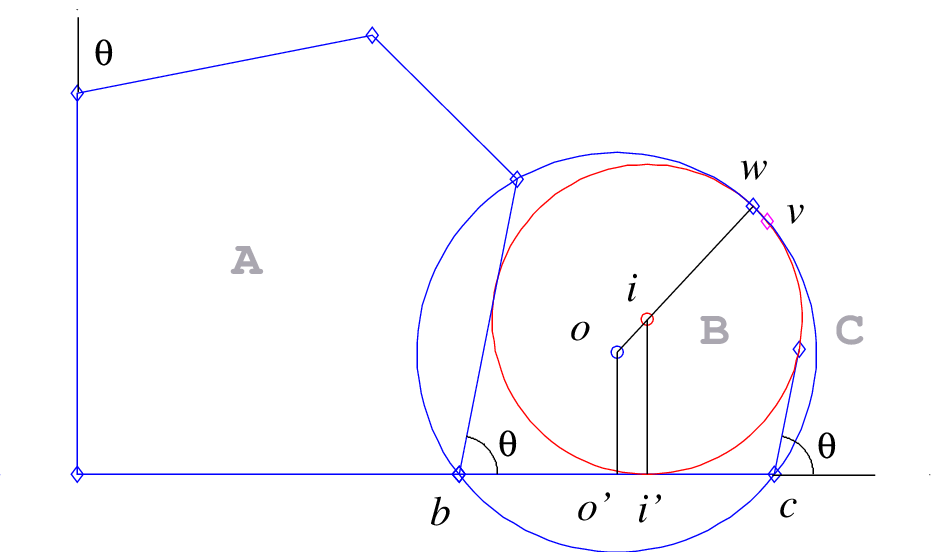} \\
(a) & (b)
\end{tabular}
}
\caption{Lower bound on $\g^*$ for a square.
(a) For $\t \approx  78.79^\circ$, $c$ falls outside $B$'s circumcircle 
(b) $B$'s circumcircle must pass through $b$, $c$ and $w$.}
\figlab{sqlb}
\end{figure}

\begin{lemma}
Pieces $A$ and $B$ of Figure~\figref{sqlb} (the corner piece and
the bottleneck piece) must satisfy
$$\min \{ \g_1(A), \g_1(B) \}  \ge 1.28868$$
\lemlab{sq.B}
\end{lemma}
\begin{pf}
Simple calculations show that for the value $\g = 1.28782$ from 
Lemma~\lemref{sq.A}, 
$B$'s circumcircle does not include point $c$ of
interior angle $\t'' \ge \t$ in $C$, as illustrated in 
Figure~\figref{sqlb}a. 
So in fact $\t$ is too large; $B$'s circumcircle is too small: 
two points ($b$, $v$) do not suffice to define $B$'s circumcircle.

Based on this observation, we force $B$'s circumcircle to pass
through three points: $b$, $c$ and the tangency point $w$ with 
the indisk (which may not lie on the bisector of $\t$).
Refer to Figure~\figref{sqlb}b. 
This leads to a new value $\g_1(B)$, which we compute next. 
To minimize $R_B$, we force $\t'' = \t$. Throughout the rest of the proof
we let the origin be determined by point $b$ and set $r_B = 1$.

Let $o = (x_o, y_o)$ and $i = (x_i, y_i)$ be $B$'s circumcenter 
and incenter, respectively. Then $y_i = 1$. 
The distance between points $b$ and $i'$ in Figure~\figref{sqlb}b 
is $1/\tan(\t/2)$. 
The distance between points $i'$ and $c$ is $1/\tan((\pi-\t)/2) =
\tan(\t/2)$. The projection $o'$ of $o$ on $bc$ falls in the middle of $bc$.
The three points $o$, $i$ and $w$ all lie on a straight line,
since $w$ is a common tangency point for the indisk and the circumcircle. 
Thus the distance between the incenter
and the circumcenter is $|oi| = R_B - 1$.
Putting all these together leads to the following system of equations:
\begin{center}
$\left\{
\begin{tabular}{l}
$x_i = 1/\tan(\t/2)$ \\
$y_i = 1$ \\
$x_o = (1/\tan(\t/2) + \tan(\t/2))/2$ \\
$x_o^2 + y_o^2 = R_B^2$ \\
$(R_B - 1)^2 = (x_i-x_o)^2 + (y_i - y_o)^2$
\end{tabular}
\right. $
\end{center}
Solving this for $y_o$ yields $y_o = (4 x_o^2 - 1)/4$. From $y_o$ we
compute $R_B = \sqrt{x_o^2 + y_o^2}$ and $\g_1(B) = 1/R_B = F(\t)$, 
where $F(\t)$ is a complex expression on $\t$.
We balance this $\g_1(B)$ with $\g_1(A)$ from (\ref{eq:ga}) and solve for $\t$. 
This yields\footnote{Mathematica
finds an exact expression for the solution, but we only report its numerical
value here.} 
$\t \approx 78.87^{\circ}$. From $\t$ we compute 
$\g_1(A) = \g_1(B) \approx 1.28868$, which completes the proof.
\end{pf}

\begin{lemma}
Any convex partition of a
a square satisfies
$\g^* > 1.28868$, i.e., the lower bound from Lemma~\lemref{sq.C} cannot be achieved.
\lemlab{sq.C}
\end{lemma}
\begin{pf}
The lower bound $\g^{*} \approx 1.28868$ derives from square
side piece $B$ with interior angles adjacent to the square side equal 
to $\t$ and $\pi - \t$. 
Now consider the sequence of pieces along the bottom square side. At some 
point the pieces in this sequence have to straighten up as they 
approach the right corner of the square. This idea is illustrated in 
Figure~\figref{halfsquare}, in which angles adjacent to the horizontal
square side have values, from left to right
$$90^\circ , 78.9^\circ , 79.0^\circ , 79.3^\circ , 80.6^\circ , 84.4^\circ , 90^\circ$$
Such straightening implies that there is a piece with interior angles
adjacent to the bottom square side equal to $\t$ and $\pi - \t'$,
with $\t' > \t$. Such a piece has ratio strictly greater than
$\g^{*}$.
\end{pf}

\begin{figure}[htbp]
\centering
\includegraphics[width=0.65\linewidth]{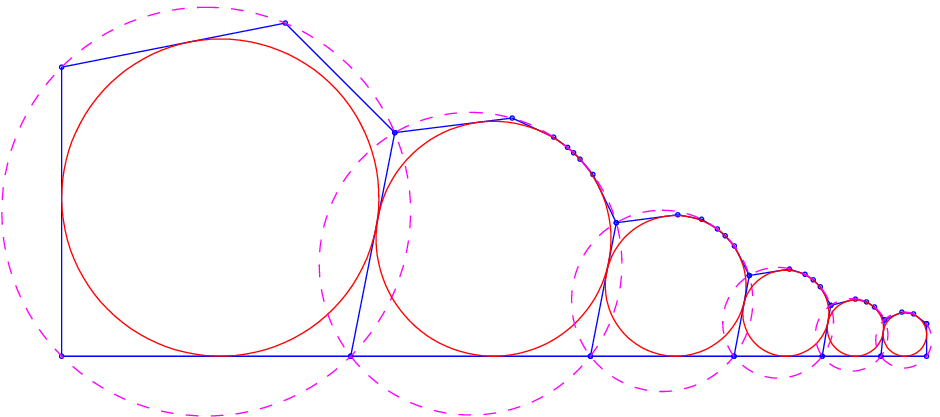}
\caption{Coverage of side of square by pieces with $\g=1.28898$.}
\figlab{halfsquare}
\end{figure}

\noindent
This completes the proof of Theorem~\theoref{square.lower.bound}.

Figure~\figref{fullsquare}a shows a coverage of the full square boundary
that achieves $\g = 1.28898$, only $3 \times 10^{-4}$ above the optimal
$\g^* = 1.28868$. This gap is determined by the bottleneck 
piece shown in Figure~\figref{fullsquare}b. 
To narrow the gap further, $\t$ and $\t'$ must approach the optimal angle 
value $78.87^\circ$ established in Theorem~\theoref{square.lower.bound},
and more points must be added in the vicinity of the tangency point 
between the indisk and the circumcircle.
This shows that the lower bound can be approached as the number of pieces
goes to infinity, but only covering the square boundary. 
As the number of pieces along the square 
sides increases, it becomes less clear how to fill the 
interior with the same ratio.

\begin{figure}[htbp]
\centering
\begin{tabular}{cc}
\includegraphics[width=0.5\linewidth]{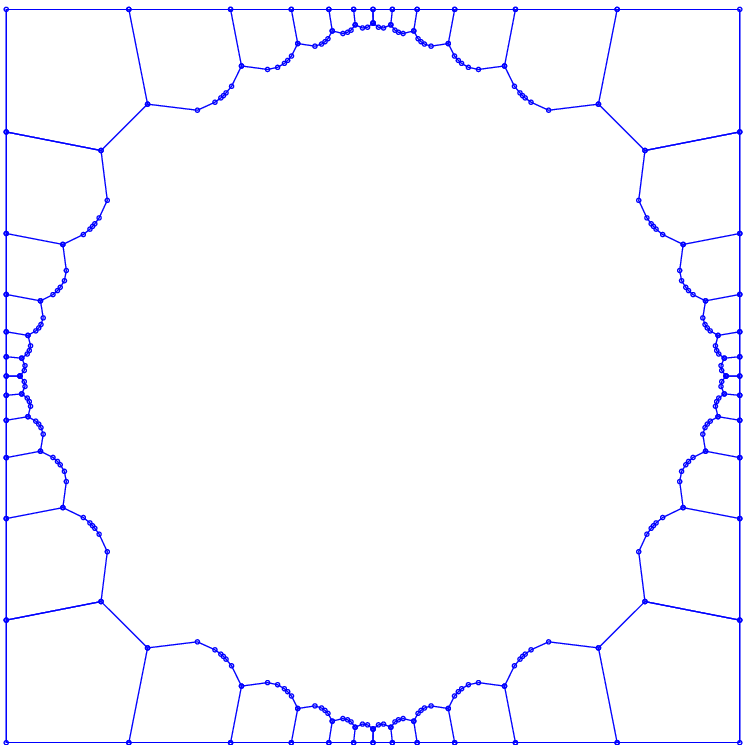} &
\includegraphics[width=0.4\linewidth]{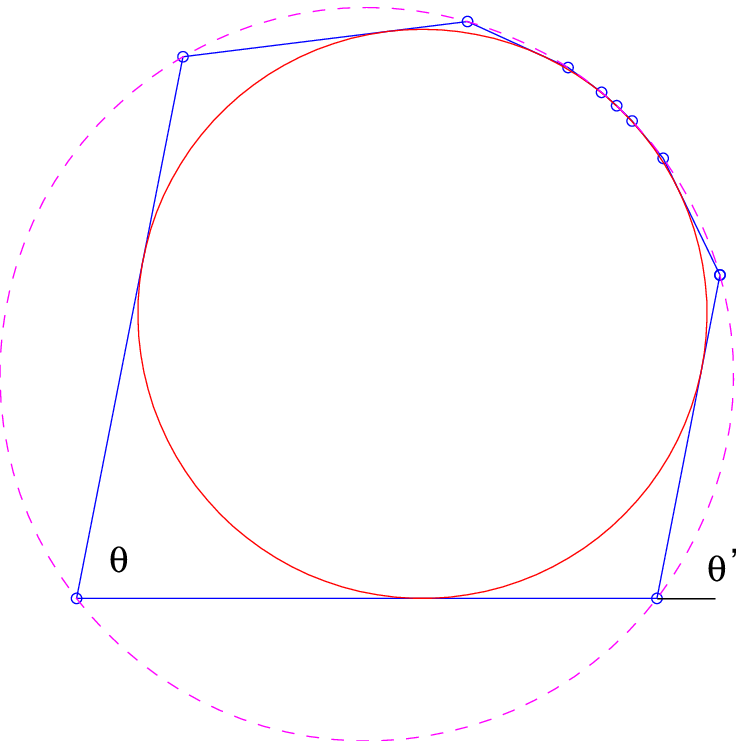}\\
(a) & (b) 
\end{tabular}
\caption{$1.28898$-coverage of full square boundary, based on Figure~\figref{halfsquare}.
(a) Overall design. (b) The ``bottleneck'' piece that establishes the lower bound.
Here $\t = 78.9^\circ$ and $\t' = 79.0^\circ$.}
\figlab{fullsquare}
\end{figure}

\subsection{Tangent Indisks}
\seclab{Tangent.Indisks}
In this section we present 
some evidence that the lower bound established in 
Theorem~\theoref{square.lower.bound}
might not be attained.
We specialize the discussion to the situation where all
the indisks touching the square side are tangent to their left
and right neighbors.  Although this is a very natural constraint,
notice in
Figure~\figref{sqlb0} that balancing the $\g$-ratios for pieces $A$
and $B$ leads to indisks that are not mutually tangent.
Of course each is tangent to the edge shared between $A$ and $B$,
but those tangency points do not coincide exactly.
The advantage of considering tangent indisks is that this added bit
of structural constraint allows some analytical calculations, rather
than relying on numerical optimizations.
Because these investigations are inconclusive, we include no proofs.
The end result, in 
Conjecture~\lemref{square.lower.bound.tangent} below, 
is a larger lower bound
for tangent indisks.

First, we compute the $\g$ for an indisk sandwiched between
parallel tangents:
\begin{lemma}
Let an indisk of radius $r$ sit on a horizontal line $L$, supported on
the left and right by tangents that both meet $L$ in angle $\t$.
Then the smallest circumcircle that includes the indisk, and the
foot of both tangents, has radius $\g r$, with
$$
\g =
        \frac{1}{4} + \csc^2(\t)
$$
The minimum value is achieved when $\t=90^\circ$, when
$\g = 5/4 = 1.25$.
\end{lemma}

\begin{figure}[htbp]
\centering
\includegraphics[width=0.4\linewidth]{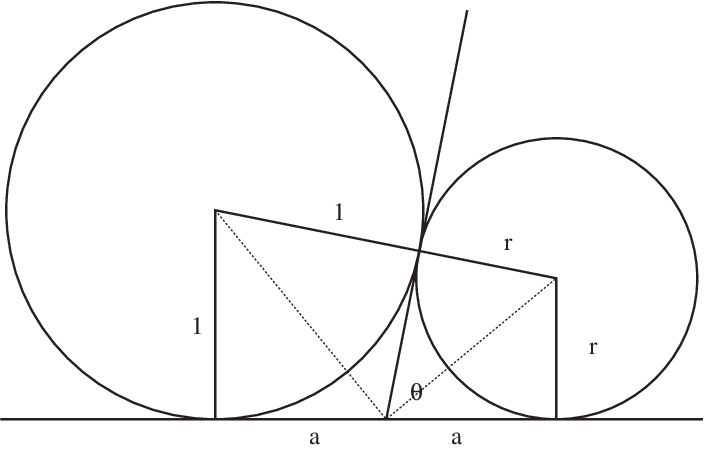}
\caption{The radii of two touching disks determines $\t$,
the angle between the tangent and the horizontal.
Here $\t=80^\circ$.}
\figlab{indisks.touching}
\end{figure}

Next we derive a relationship between the angle of the separating
tangent line and the disk radii;
see Figure~\figref{indisks.touching}.
\begin{lemma}
Let two disks, of radius $1$ and $r < 1$, touch a horizontal line $L$,
and touch each other.
Then the line $T$ of tangency between them
is slanted toward the smaller disk, making an angle $\t$
with the horizontal, and meeting $L$ midway between the
projection of the centers
of the two disks, with
$$
r = \tan^2 ( \t/2 )
$$
\end{lemma}

\begin{figure}[htbp]
\centering
\includegraphics[width=0.5\linewidth]{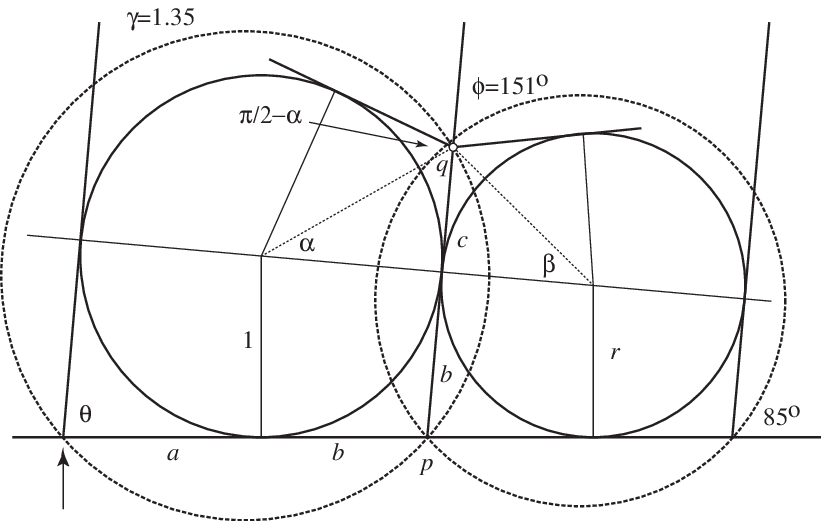}
\caption{$q$, and therefore $\phi$, are determined by $\t$ and $\g$.}
\figlab{slanted.2}
\end{figure}

Third, we compute the angle $\phi$ at the next ``level'' between two
adjacent pieces, as a function of the tangent slant $\t$ and $\g$;
see Figure~\figref{slanted.2}.
\begin{lemma}
Let two tangent indisks, of radius $1$ and $r$, sit on a horizontal line $L$, supported
by three slanted tangents that all
meet $L$ in angle $\t$.
Let the origin be determined by the leftmost tangent's intersection with $L$.
Let $T$ be the middle tangent, with $T \cap L = p =(p_x,p_y)$.
Let circumcircles of radius $\g$ and $\g r$ surround the indisks and pass
through the feet of the tangents on $L$.
Let $q=(q_x,q_y)$ be the point on $T$ that is highest above $L$
and inside both circumcircles and $\phi$ the 
angle determined by $q$ and the upper tangents to
the two disks.
Then $\phi$, as a function of $\t$ and $\g$, is
$$\phi/2 =
\tan^{-1} [
   \cot( \t/2 )
   s ]
+
\tan^{-1} [
   2 s ]
$$
where
$$
s = \sqrt{
-1
+ \cot^2 \t
-2 \cos \t
\sqrt{ \g^2 - \csc^2 \t}
+ \g^2 \sin^2 \t
}
$$
\end{lemma}

\begin{figure}[htbp]
\centering
\includegraphics[width=0.6\linewidth]{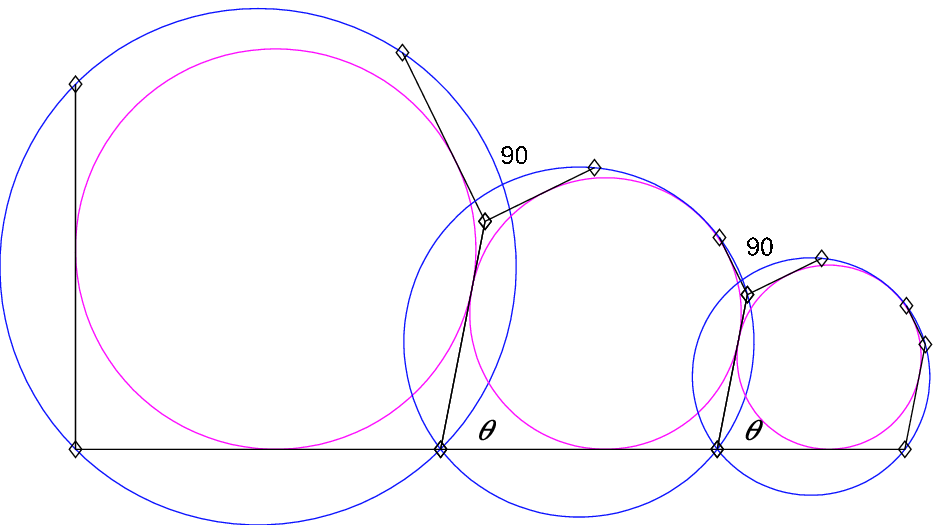}
\caption{Lower bound for tangent indisks $\g^* = 1.28939$. Here $\t =
78.94^\circ$.}
\figlab{square_lb_tangent}
\end{figure}

Finally, the above three lemmas permit us to start with a $\phi$ of
$90^\circ$ and compute backwards. This leads to a lower bound in the ``tangent indisks'' model where we tune $\gamma$ so that the next-level angle $\phi$ is $90^\circ$. 
See Figure~\figref{square_lb_tangent}.

\begin{conj}
The optimal aspect ratio for a partition of a square into convex pieces with
tangent indisks along the sides of the square is 
$\g^* \ge  1.28939$.
\lemlab{square.lower.bound.tangent}
\end{conj}

\noindent
The idea here is that with this $\g$, we have duplicated the corner of
the square at the next level, leaving the remaining problem of covering
the interior as difficult as the original. 

\subsection{Square: Discussion}
\seclab{Square.Open}
Results of Sections~\secref{Pentagons.on.Side}--\secref{Square.Lower.Bound} 
show that finding an optimal aspect 
ratio partition of a square is not straightforward.
It remains open to narrow the $0.01082$ gap between the lower bound of $1.28868$ 
provided by Lemma~\theoref{square.lower.bound} and the upper bound
of $1.29950$ established by Figure~\figref{stainedglass}.

\begin{figure}[htbp]
\centering
\begin{tabular}{cc}
\includegraphics[width=0.48\linewidth]{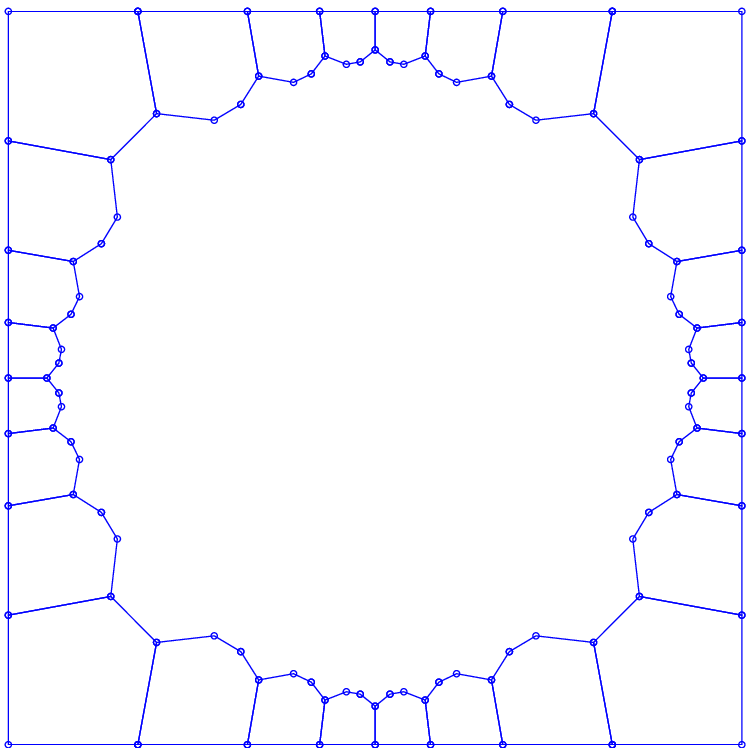} &
\includegraphics[width=0.45\linewidth]{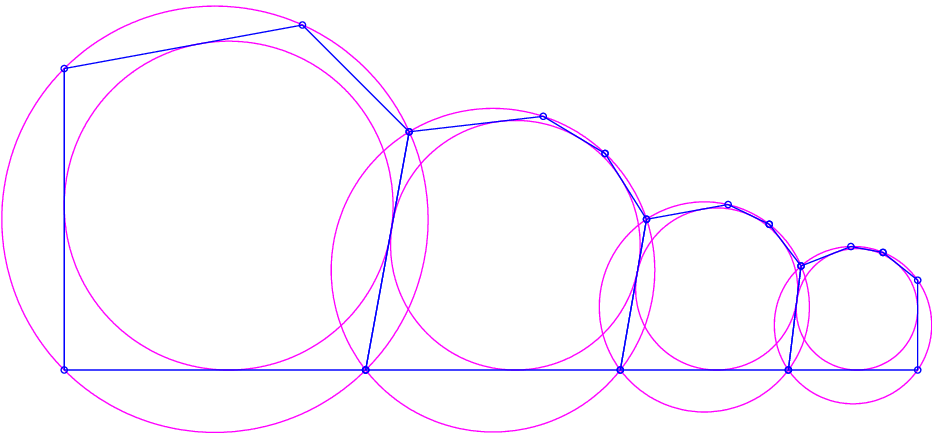} \\
(a) & (b)
\end{tabular}
\caption{Partial partition with $\g = 1.29650$. (a) Coverage of full
boundary. (b) Details of critical indisks and circumcircles.}
\figlab{square6}
\end{figure}

As mentioned, it seems feasible to lower the upper bound of the gap to 
a value slightly larger than $\g = 1.29625$, the optimal $\g$ value for 
any partition that covers the entire square boundary with hexagons.
For instance, Figure~\figref{square6} shows a coverage of the full 
square boundary that achieves $\g=1.29650$, which is only $0.00025$ 
above the optimal for hexagons. 
We conjecture that a construction similar to that in
Figure~\figref{stainedglass} can be used to fill in the interior.

To narrow the gap further, it is necessary to employ pieces with more 
than six vertices along the boundary of the square. 
However, Figure~\figref{fullsquare}
suggests that pieces with a large number of vertices along square sides 
create a large discrepancy in segment lengths bounding the interior.  
Filling the interior then becomes problematic. 
However we see no fundamental impediment to doing so. 
Based on partial results not reported in this paper, we would be
surprised if the optimal partition could be achieved with a finite partition:

\begin{conj}
No finite partition achieves the optimal partition of the square:
rather $\g^*$ can be approached as closely as desired as
the number of pieces goes to infinity.
\end{conj}

\notyet{
We have investigated
a few sophisticated techniques for filling the interior 
of Figure~\figref{fullsquare}, and based 
on some partial results not reported in this paper, we conjecture the
following:
\begin{conj}
A square may be partitioned into a finite number of pieces with ratio 
$\g$, for any $\g > 1.28868$.
As $\g$ approaches $1.28868$, the number of pieces goes to infinity.
\end{conj}
}

\section{Discussion}
\seclab{Discussion}
This paper establishes the optimally circular convex 
partition of all regular polygons, except for the square,
where we have left a small gap.
We hope to show in future work that our results apply to
arbitrary polygons as well.
We have also investigated nonconvex circular partitions of
regular polygons~\cite{do-prpcp2-03}.

Our work leaves many problems unresolved:
\begin{enumerate}
\item Narrow
the gap $[1.28868, 1.29950]$ for the 
optimal aspect ratio of a convex partition of a square.
\item Determine for the square if $k^*$, the number of pieces
in an optimal partition, is finite or infinite
(cf.~Conjecture~1).
\item For each $k$, find the optimal ratio $\g^{(k)}$ of a polygon $P$ using
only convex $(\le k)$-gons.
It is especially interesting to determine $\g^{(k+1)}$ for a $k$-gon.
For example, Figure~\figref{Pentagonal} and work
not reported here shows that $\g^{(5)} \in [1.31408, 1.33964]$
for a square.
We did not explore $\g^{(4)}$ for an equilateral triangle;
it would be interesting to replace
the corner pieces in Figure~\figref{eqtri.all} with quadrilaterals.
\item For each $k$, find the optimal ratio $\g_k$ of a polygon $P$ using
$\le k$ convex pieces.
So each of our partitions of the square establishes a particular upper bound,
e.g.,  Figure~\figref{sq.4.convex} shows that $\g_{21} \le 1.32348$.
\item More generally, develop a tradeoff between the number of pieces of the partition
and the circularity ratio achieved.
\item Extend our results to all convex polygons, and then to arbitrary polygons.
\end{enumerate}
Finally, all these problems could be fruitfully explored in 3D, the natural 
dimension for the applications discussed in Section~\secref{Motivation}. 

\bibliographystyle{alpha}
\bibliography{Convex}

\end{document}

%% file: mac.tex
\newenvironment{pf}{\unskip{\bf Proof:}}{\unskip{\hfill $\Box$}}

\newcommand{\lemlab}[1]{\label{lemma:#1}}
\newcommand{\theolab}[1]{\label{theo:#1}}

\newcommand{\tablab}[1]{\label{tab:#1}}
\newcommand{\figlab}[1]{\label{fig:#1}}
\newcommand{\seclab}[1]{\label{section:#1}}

\newcommand{\lemref}[1]{\ref{lemma:#1}}

\newcommand{\theoref}[1]{\ref{theo:#1}}

\newcommand{\tabref}[1]{\ref{tab:#1}}
\newcommand{\figref}[1]{\ref{fig:#1}}
\newcommand{\eqref}[1]{(\ref{eq:#1})}
\newcommand{\secref}[1]{\ref{section:#1}}

\newtheorem{theorem}{Theorem} 
\newtheorem{lemma}[theorem]{Lemma}

\newtheorem{conj}{Conjecture} 
{\catcode`\@=11
\gdef\setft#1#2#3{%
\def\@oddfoot{
{\setbox0=\hbox{#1}
\setbox1=\hbox{#3}
\ifdim\wd0>\wd1
\dimen0=\wd0
\box0\hfil#2\hfil\hbox to\dimen0{\hfil\hfil\box1}
\else \dimen0=\wd1
\hbox to\dimen0{\box0\hfil }\hfil#2\hfil\box1 \fi
}}} }

\def\complaint#1{}
\def\withcomplaints{
\newcounter{mycomplaints}
\def\complaint##1{\refstepcounter{mycomplaints}%
\ifhmode%
\unskip%
{\dimen1=\baselineskip \divide\dimen1 by 2 %
\raise\dimen1\llap{\tiny -\themycomplaints-}}\fi%
\marginpar{\tiny [\themycomplaints]: ##1}}%
}